\documentclass[acmsmall,screen]{acmart}

\usepackage{xspace}
\usepackage{threeparttable}
\usepackage{graphicx}
\usepackage{caption}
\usepackage{textcomp}
\usepackage{framed}
\usepackage{url}

\usepackage{breakurl}
\usepackage{xcolor}
\usepackage{multirow}
\usepackage{tabularx}

\usepackage{enumitem}
\usepackage{subfigure}
\usepackage[linesnumbered,ruled,vlined]{algorithm2e}
\usepackage{tcolorbox}
\usepackage{threeparttable}
\usepackage{pifont}
\usepackage{wasysym}

\usepackage[nameinlink]{cleveref}
\crefname{figure}{Figure}{Figures}
\crefname{table}{Table}{Tables}
\crefname{algorithm}{Algorithm}{Algorithms}
\crefname{section}{Section}{Sections}
\definecolor{acmcolor}{RGB}{108, 35, 130}
\definecolor{UTG}{RGB}{8, 94, 69}
\definecolor{CCFG}{RGB}{112, 48, 160}
\definecolor{UDFG}{RGB}{230, 60, 94}

\newcommand{\tool}{\textsc{MiniScope}\xspace}

\begin{document}

% \title{\tool: Detecting Privacy Inconsistency of MiniApps via Dual-round Hybrid Analysis}
\title{\tool: Automated UI Exploration and Privacy Inconsistency Detection of MiniApps via Two-phase Iterative Hybrid Analysis}

\author{Shenao Wang}
\authornote{Hubei Key Laboratory of Distributed System Security, Hubei Engineering Research Center on Big Data Security, School of Cyber Science and Engineering, Huazhong University of Science and Technology.}
\email{shenaowang@hust.edu.cn}
\orcid{0000-0003-3818-3343}
\affiliation{%
  \institution{Huazhong University of Science and Technology}
  \city{Wuhan}           
  \country{China}
}

\author{Yuekang Li}
\email{yli044@e.ntu.edu.sg}
\orcid{0000-0003-4382-0757}
\affiliation{%
  \institution{University of New South Wales}
  \city{Sydney}
  \country{Australia}
}

\author{Kailong Wang}
\authornotemark[1]
\authornotemark[2]
\email{wangkl@hust.edu.cn}
\orcid{0000-0002-3977-6573}
\affiliation{%
  \institution{Huazhong University of Science and Technology}
  \city{Wuhan}
  \country{China}
}

\author{Yi Liu}
\email{yi009@e.ntu.edu.sg}
\orcid{0000-0002-4978-127X}
\affiliation{%
  \institution{Nanyang Technological University}
  \country{Singapore}
}

\author{Hui Li}
\email{lihui@mail.xidian.edu.cn}
\orcid{0000-0001-8310-7169}
\affiliation{
    \institution{Xidian University}
    \city{Xi'an}
    \country{China}
}

\author{Yang Liu}
\email{yangliu@ntu.edu.sg}
\orcid{0000-0001-7300-9215}
\affiliation{
    \institution{Nanyang Technological University}
    \country{Singapore}
}

\author{Haoyu Wang}
\authornotemark[1]
\authornote{Haoyu Wang (haoyuwang@hust.edu.cn) and Kailong Wang (wangkl@hust.edu.cn) are the corresponding authors.}
\email{haoyuwang@hust.edu.cn}
\orcid{0000-0003-1100-8633}
\affiliation{%
  \institution{Huazhong University of Science and Technology}
  \city{Wuhan}     
  \country{China}
}

\renewcommand{\shortauthors}{Wang et al.}

\begin{abstract}
  The advent of MiniApps, operating within larger SuperApps, has revolutionized user experiences by offering a wide range of services without the need for individual app downloads. However, this convenience has raised significant privacy concerns, as these MiniApps often require access to sensitive data, potentially leading to privacy violations. Despite existing privacy regulations and platform guidelines, there is a lack of effective mechanisms to safeguard user privacy fully.
  To address this critical gap, we introduce \tool, a novel two-phase hybrid analysis approach, specifically designed for the MiniApp environment. This approach overcomes the limitations of existing static analysis techniques by incorporating UI transition states analysis, cross-package callback control flow resolution, and automated iterative UI exploration. This allows for a comprehensive understanding of MiniApps' privacy practices, addressing the unique challenges of sub-package loading and event-driven callbacks. Our empirical evaluation of over 120K MiniApps using \tool demonstrates its effectiveness in identifying privacy inconsistencies. The results reveal significant issues, with 5.7\% of MiniApps over-collecting private data and 33.4\% overclaiming data collection. We have responsibly disclosed our findings to 2,282 developers, receiving 44 acknowledgments. These findings emphasize the urgent need for more precise privacy monitoring systems and highlight the responsibility of SuperApp operators to enforce stricter privacy measures. 
\end{abstract}




\maketitle

\section{Introduction}
Mini-programs or MiniApps\footnote{Considering the openness of the Android ecosystem and numerous prior studies focusing on WeChat MiniApps~\cite{wang2023taintmini,wang2023doasyousay,zhang2023spochecker}, our work particularly focuses on the same target, i.e., the WeChat MiniApp in the Android, unless stated otherwise.}, epitomizing a novel breed of lightweight mobile applications, have been transformative in the realm of user experience in recent years~\cite{yang2022cross}. 
MiniApps are lightweight versions of full-fledged applications that operate within a host application or ``SuperApp''.
They can provide users with a diverse range of services under the ``SuperApp + MiniApps'' business paradigm umbrella~\cite{Kumar_2018}. 
From online shopping, gaming, and accessing social media to availing healthcare services~\cite{wechatshopping, rao2021impulsive, hao2018analysis}, these MiniApps offer the richness of standalone applications without necessitating individual downloads. 
This seamless integration has radically enhanced user experiences, with simplified navigation and easy-to-access features all within a single app interface. 
For instance, users can effortlessly switch between chatting with friends, making payments, ordering food, scheduling medical appointments, streaming multimedia content, and even accessing government services—all within a single SuperApp.
Currently, the most representative SuperApp, WeChat, hosts a staggering 3.5 million MiniApps that engage over 600 million daily active users~\cite{yang2023sok}. 
Its extensive ecosystem underscores the revolutionary impact of the MiniApp paradigm, which could introduce complications and unexpected privacy concerns at the same time.

To deliver their diverse services, these MiniApps often require access to sensitive system resources~(e.g., camera and Bluetooth), as well as user data~(e.g., location information, phone numbers, and email accounts). 
This has given rise to potential privacy violations, as third-party developers within the SuperApp ecosystem could access and utilize this information, often without clear or explicit user consent. 
Accompanied by the rising awareness and concern for user privacy, numerous regulations and laws, including the General Data Protection Regulation (GDPR)~\cite{GDPR}, California Consumer Privacy Act (CCPA)~\cite{CCPA}, Act on the Protection of Personal Information (APPI)~\cite{APPI}, and Canadian Consumer Privacy Protection Act (CPPA)~\cite{CPPA}, have been enforced to ensure the transparent and accountable collection and processing of user data. 
Meanwhile, SuperApp platforms, including WeChat, have also incorporated guidelines that demand stricter scrutiny of data collection and processing from MiniApps~\cite{Wechat_privacypolicy}. 
However, these measures, while laudable, have proven insufficient to fully safeguard user privacy within the expansive SuperApp ecosystems~\cite{Wechat_risk2,Wechat_risk1}. 

\noindent \textbf{Research Gaps.}
While considerable progress has been made in the field of privacy inconsistency analysis for mobile apps~\cite{slavin2016toward,wang2018guileak,zimmeck2016automated} and Web apps~\cite{ling2022arethey,bui2023detection}, the unique features of MiniApps present distinct challenges that remain largely unexplored. 
Existing techniques, primarily centered around taint analysis~\cite{arzt2014flowdroid,enck2014taintdroid,wang2023taintmini,meng2023wemint,li2023minitracker}, encounter two major limitations when adapted to the MiniApp environment.
Firstly, unlike traditional app packaging mechanisms that allows access to all source code at once, MiniApps employ a subpackaging mechanism~\cite{subpackaging}, that is, only the main package is initialized during a cold start and the sub-packages are loaded on demand during runtime. Consequently, methods that rely solely on static analysis~\cite{wang2023taintmini,meng2023wemint,li2023minitracker} fail to trigger the loading of these sub-packages, thereby overlooking a substantial amount of code that could involve privacy practices. This oversight leads to an incomplete data flow graph, resulting in a higher incidence of false negatives in taint analysis.
Secondly, MiniApps are framework-based and event-driven. Prior research~\cite{yang2015ccfg,liu2022promal,liu2023ex} has demonstrated the critical importance of user-driven callback control flow analysis in similar application paradigm. However, existing taint analysis methods~\cite{wang2023taintmini,meng2023wemint,li2023minitracker} concentrate primarily on sensitive data flows while neglecting event-driven callbacks, falling short in thoroughly scrutinizing asynchronous entry points of MiniApp logics. This neglect results in a significant number of false positives in taint analysis, as some taint paths obtained from unused functions and orphaned pages could be potentially unreachable.

\noindent \textbf{Technical Challenges.} 
Given the above research gaps, our goal is to develop a framework that encompasses subpackage loading and integrates callback control flow analysis. To achieve this, we face three key technical challenges. 
1) Firstly, the static analysis is intricately intertwined with dynamic UI exploration. That is, in the context of MiniApps, the prerequisite for sound static analysis is to load sub-packages on demand through UI exploration, while efficient UI exploration requires static prior knowledge (such as UI transition states) for guidance. Therefore, this presents a logical paradox, which demands designing tailored hybrid analysis strategies specifically for MiniApps.
2) Secondly, for effectively guided UI exploration, we need to construct a comprehensive UI state transition graph. However, dynamic data binding and page routing create complex interactions between JavaScript and WXML. Unlike Android where transitions are often explicitly defined through Intent objects and are relatively straightforward to trace, MiniApp involves a more intricate process due to the dynamic nature of data binding and cross-language interactions. Tracking and analyzing this type of UI state transition is challenging.
3) Finally, the control flow analysis of MiniApps presents difficulties due to the need for complete resolution of user interactions and event-driven callbacks.
Although similar techniques have already been widely explored in Android/iOS~\cite{yang2015ccfg,liu2022promal,liu2023ex}, the unique challenge in constructing control flow in MiniApps lies in the dynamic definition of callbacks, especially the context binding when reusing callback functions across different files, which we will further explain in \S\ref{sec:challenges}.

\noindent \textbf{Our Approach.} 
In response to the research gaps and challenges in analyzing MiniApps, we introduce \tool, an innovative two-phase iterative hybrid analysis approach tailored for in-depth UI exploration and precise privacy practices identification.
The initial phase of our approach focuses on dynamic sub-package loading through UI exploration to obtain the complete package. Utilizing a static analyzer, we initialize a MiniApp Dependency Graph (MDG) for the main package, meticulously designed to extract UI transition states, event-driven control flows, and data flows. This initial MDG serves as the groundwork for accurate taint analysis and guides the subsequent UI exploration. Based on the initial MDG, the directed UI explorer then engages in fuzzy matching between WXML components and the UI Widgets Tree. By adopting a sub-package-directed breadth-first traversal strategy, \tool dynamically explores sub-package pages, thereby obtaining the complete package of the MiniApp. In the second phase, our focus shifts to the precise identification of privacy practices. \tool merges the main package with the dynamically loaded sub-packages, performing static analysis to obtain the complete MDG. In this phase, the directed UI explorer utilizes a privacy-practice-directed depth-first strategy for runtime exploration of sensitive behaviors. \tool then cross-verifies the privacy practices extracted from both static and dynamic analysis against the declared privacy policy, enabling a thorough detection to identify any discrepancies or privacy inconsistencies.

To demonstrate the effectiveness practically, we utilize \tool to detect privacy inconsistency and carry out a comprehensive evaluation involving over 120K MiniApps.
Following a filtration process to exclude those without a valid privacy policy, we eventually analyzed 10,786 MiniApps. 
The results underscore the superior performance of \tool in identifying privacy-related practices when compared with \textsc{TaintMini}~\cite{wang2023taintmini}, the state-of-the-art technique. 
Specifically, \tool demonstrates 7.9\% and 23.5\% improvement in precision and recall respectively. 
Our analysis further reveals that 5.7\% of the MiniApps are secretly over-collecting private data, and 33.4\% of the MiniApps overclaim the data they actually collect. 
We have responsibly disclosed our findings to 2,282 developers, receiving 44 confirmations and acknowledgments. 
These findings highlight the pressing need for the implementation of a more precise privacy monitoring system, emphasizing the responsibility of SuperApp operators to enforce such measures.

\noindent \textbf{Contributions.} Our contributions are summarized as follows:
\begin{itemize}[leftmargin=2em]

\item \textbf{Novel Techniques~($\S$\ref{sec:methodology}).} 
This work introduces a novel two-phase hybrid analysis approach for MiniApps, including UI transition states modeling, detailed cross-file callback control flow resolution, and automated iterative UI exploration, bridging the research gap of sub-package loading and callback analysis in the MiniApp ecosystem.

\item \textbf{Practical Implementation and Application~($\S$\ref{sec:Privacy Compliance Detector})} 
We have implemented these techniques as a fully automatic artifact named \tool and demonstrated its performance and effectiveness for a comprehensive detection of privacy inconsistency in MiniApps.
In support of open science, we release the source code of \tool available~\cite{miniscope}.

\item \textbf{Empirical Evaluation and Real-world Impacts~($\S$\ref{sec:evaluation}).} We have conducted a comprehensive evaluation on 120K MiniApps, revealing prevalent over-collection and overclaim of privacy information among 5.7\% and 33.4\% of the MiniApps respectively. We reported our findings to over 2K developers, receiving 44 confirmations from them. 

\end{itemize}

\section{Background}

\subsection{Features of MiniApps}
\begin{figure}[!htbp]
    \centering
    \includegraphics[width=0.55\textwidth]{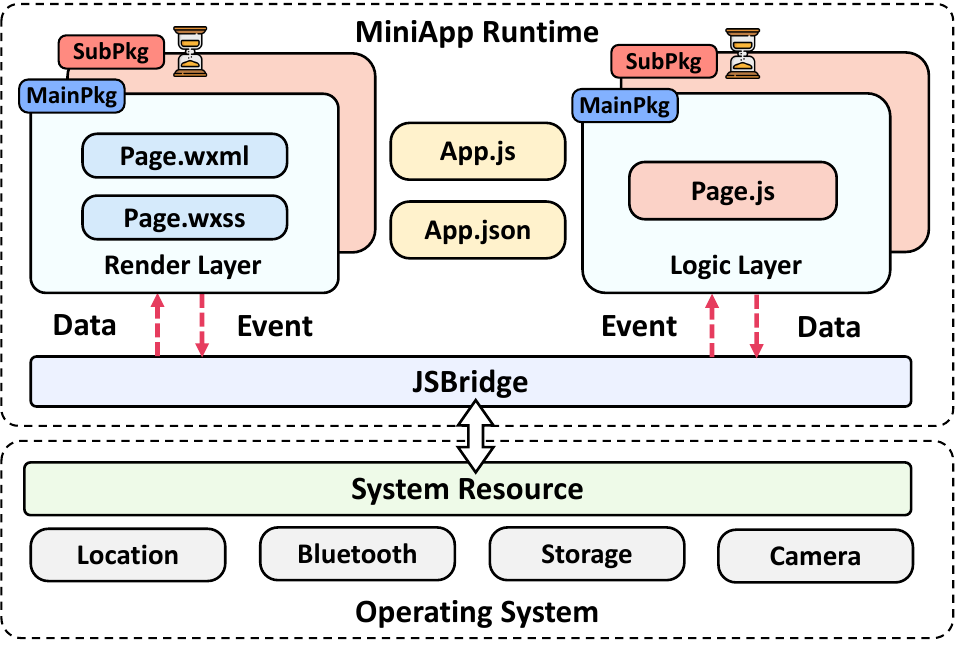}
    \caption{Architecture of MiniApps.}
    \label{fig:architecture}
\end{figure}

\noindent\textbf{Architecture of MiniApps.} The runtime environment of a MiniApp (\cref{fig:architecture}), facilitated by SuperApps like WeChat or Alipay, is characterized by a dual-thread architecture splitting into a render~(or view) layer and a logic layer~\cite{wang2023taintmini}. The render layer, akin to HTML and CSS, involves \texttt{WXML}~(Weixin Markup Language) and \texttt{WXSS}~(Weixin Style Sheet), while the logic layer consists of JavaScript files. MiniApp pages correspond to multiple WebView threads in the render layer, whereas logical operations, data requests, and interface calls occur via JSCore threads in the logic layer. The two layers engage through the JSBridge mechanism, handling user interaction events and data updates. Furthermore, the logic layer introduces sub-app APIs like \texttt{wx.getLocation} for user location retrieval, which are bridged into the SuperApp's native layer~\cite{lu2020demystifying}. 

% \noindent \textbf{Architecture of MiniApps.} 

\noindent\textbf{Sub-package Dynamic Loading.}
MiniApp employs a unique sub-package mechanism~\cite{w3c} to stay lightweight. During a cold-start or initialization, only the main package is loaded, with corresponding sub-packages loaded dynamically as users access specific pages. Subpackage invocation typically involves specifying the subpackage page path in the configuration file~(e.g., app.json in WeChat). The platform then handles downloading the specified subpackage when a user navigates to a page that requires it. 

\noindent\textbf{Page Routing of MiniApps.}
WeChat MiniApps are multi-functional entities, with each page facilitating specific tasks and interconnected via page routing.
Once initiated, the WeChat client establishes a page stack, enabling page manipulations for the MiniApp. Two methods facilitate this routing process: the \texttt{<navigator>} widget in the render layer and the navigation-specific sub-app APIs in the logic layer. The former offers both intra-MiniApp and cross-MiniApp navigation, governed by the widget's \texttt{target} attribute, with a variety of routing methods based on the \texttt{open-type} attribute. The latter implements routing in the logic layer, using the \texttt{url} property to specify the target page, with APIs like \texttt{wx.navigateTo} and \texttt{wx.redirectTo} enabling different navigational outcomes.

\noindent\textbf{Data Binding.}
Data binding in MiniApps allows developers to synchronize the UI with the application data model. This is achieved using a combination of WXML and JavaScript. Developers can bind data to UI components using the \{\{\}\} syntax in WXML, which automatically updates the UI whenever the corresponding JavaScript data changes. For example, the expression \texttt{<view>\{\{message\}\}<\/view>} binds the \texttt{message} variable from the JavaScript context to the view component, ensuring that any changes to \texttt{message} are reflected in the UI.

\noindent\textbf{Event-driven Callbacks.}
MiniApps, being event-driven, respond to UI interactions with appropriate logic execution and interface updates. Consequently, static analysis of MiniApps should consider the UI event-driven callback functions that influence the control flow. 
This work focuses on two following crucial types of callbacks: 

\noindent $\bullet$ \textbf{Lifecycle Callbacks} include two-pronged MiniApp lifecycle—\texttt{App} instance and \texttt{Page} instance lifecycles. For instance, a cold-started MiniApp first triggers the \texttt{onLaunch} callback of the \texttt{App} instance, and when the MiniApp launches and surfaces, it triggers the \texttt{onShow} callback. A page's first load activates the \texttt{onLoad} callback of the \texttt{Page} instance, which in turn triggers the \texttt{onShow} callback when displayed and stacked.

\noindent $\bullet$ \textbf{GUI Event Handler Callbacks~(Event Binding)} facilitate communication between a MiniApp's rendering and logic layers. This is done using attributes like \texttt{bindtap} in WXML, which specifies a function to be called when the event occurs. For instance, \texttt{<button bindtap=``handleTap''>Click me<\/button>} binds the \texttt{handleTap} function to the button's \texttt{tap} event. When the user taps the button, the \texttt{handleTap} function is invoked to respond to user actions.

\subsection{Comparison with Native/Web Apps}
In contemporary application development, MiniApps, Native Apps, and Web Apps represent three distinct paradigms. Although there are similarities among them, each one possesses its unique features and functionalities. In the following, we elucidate the differential features of these platforms, particularly focusing on two primary aspects: 

\begin{table}[!htbp]
\centering
\begin{threeparttable}
\caption{Comparison of MiniApps~(WeChat), Native Apps, and Web Apps. }
\fontsize{8}{12}\selectfont
\begin{tabular}{cccc}
\toprule
\textbf{Mechanism} & \textbf{MiniApps~(WeChat)} & \textbf{Native Apps} & \textbf{Web Apps} \\
\hline
\multirow{2}{*}{\textbf{Packaging Format}} & \multirow{2}{*}{WXAPKG} & APK (Android) & \multirow{2}{*}{/} \\ 
 & & IPA (iOS) & \\
\hline
\textbf{Building Mechanism} & Subpacking & Compilation \& Packaging & Bundling \\ 
\hline
\multirow{2}{*}{\textbf{Rendering Mechanism}} & Hybrid Rendering  & \multirow{2}{*}{Native Rendering} & \multirow{2}{*}{HTML/CSS Rendering} \\ 
 & (WebView \& Native) & & \\
\hline
\multirow{2}{*}{\textbf{Layout Code}} & \multirow{2}{*}{WXML \& WXSS} & XML~(Android)  & \multirow{2}{*}{HTML \& CSS} \\ 
 & & Storyboard \& SwiftUI (iOS) & \\
\hline
\multirow{2}{*}{\textbf{Logic Code}} & \multirow{2}{*}{JavaScript} & Java (Android) & \multirow{2}{*}{JavaScript} \\ 
 & & Swift or Objective-C (iOS) & \\
\hline
\multirow{2}{*}{\textbf{Page Routing}} & Platform API~(e.g. wx.navigateTo) & Intent~(Android) & \multirow{2}{*}{Tag <a>}  \\
 & Tag <navigator> & SwiftUI Navigation~(iOS) & \\
\hline
\textbf{Distribution} & SuperApp~(e.g., WeChat) & App Store~(e.g., Google Play) & Website Access \\ 
\bottomrule
\end{tabular}
\end{threeparttable}
\label{tab:comparison}
\end{table}

\noindent\textbf{Unique Packaging and Building Mechanism:} Unlike web apps that lack a specific packaging format, MiniApps, and native apps require specific file formats for distribution and installation. MiniApps, commonly packaged as WXAPKG, diverge notably from native apps, which utilize APK or IPA formats for Android and iOS platforms, respectively. 
The building process of these applications also presents significant differences. 
Unlike native apps where the packaging process often allows access to all source code at once, the subpacking mechanism in MiniApps inherently limits the static acquisition of code to only the main package. Subpackage code in MiniApps is not immediately available; it requires dynamic traversal and is loaded as needed, a stark difference from the comprehensive code availability seen in native app bundling.

\noindent\textbf{Hybrid Rendering Mechanism:}  
MiniApps leverage a hybrid rendering approach specifically designed for the WeChat ecosystem, combining WebView and Native components. 
Its unique components, such as \texttt{<navigator>}, akin to a hybrid of HTML's \texttt{<a>} tag and native app navigation functions, exemplify integration with WeChat's user experience. 
Media components like \texttt{<image>}, \texttt{<video>}, and \texttt{<camera>} in WXML, while functionally similar to their counterparts in Native and Web Apps, are fine-tuned for performance and direct integration with WeChat platform. 

\section{Motivating Example}
\begin{figure}[!htbp]
    \centering
    \includegraphics[width=0.98\linewidth]{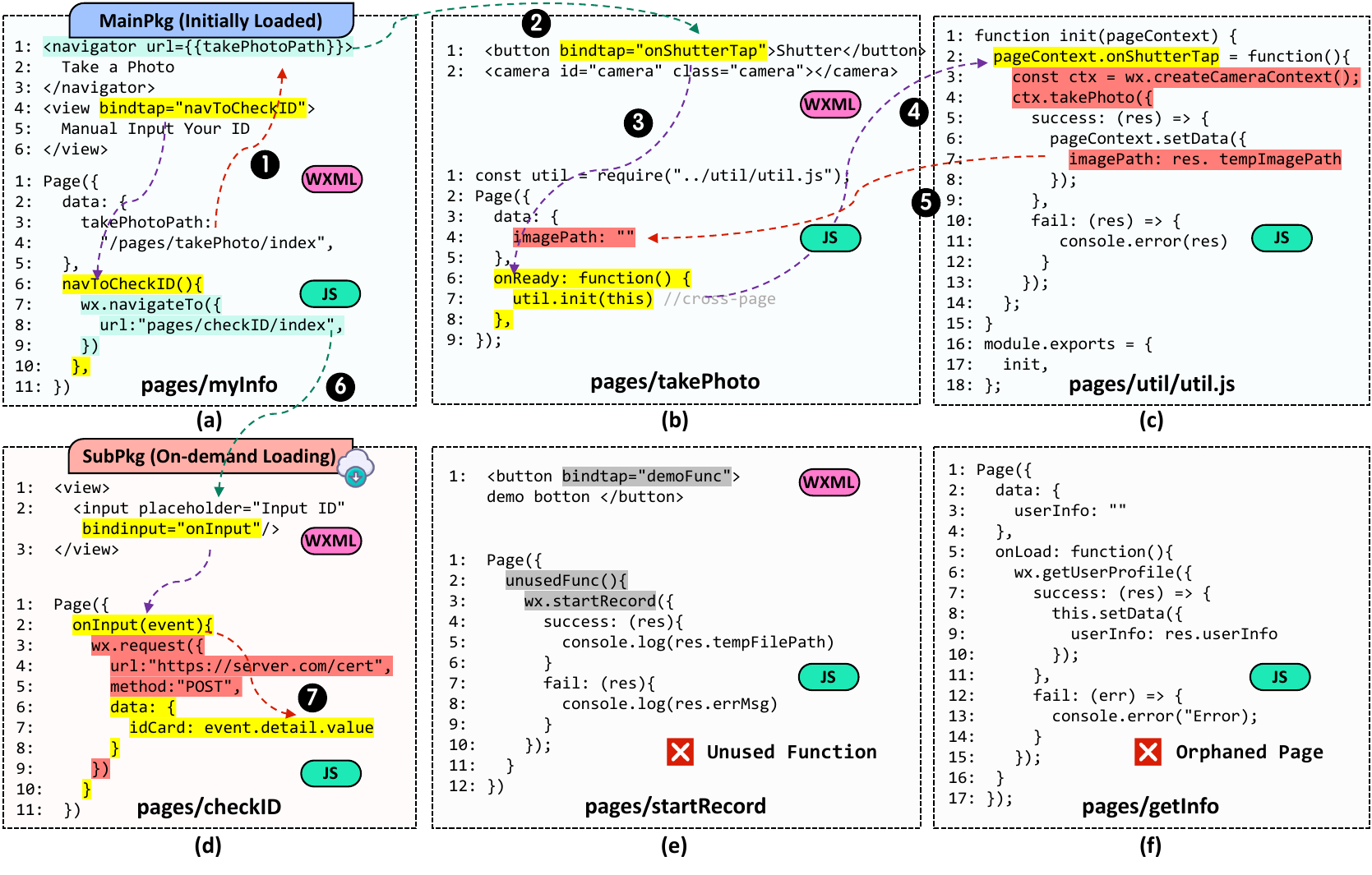}
    \caption{Code snippet of cross-file/cross-package privacy practices. Subfigure (a), (b), and (c) illustrate cross-file privacy practices in the initially loaded main package, subfigure (a) and (d) represent cross-subpackage privacy practices (i.e., FNs in previous static-only approaches, but identified by \tool), and subfigure (e) and (f) demonstrates dead code~(unused function and orphaned page, respectively) without entry points (i.e., FPs generated in previous DFA-only approaches, but excluded by \tool). }
    \label{fig:Code Snippet}
\end{figure}

To illustrate the challenges of identifying the privacy practice of MiniApps, we provide a running example in \cref{fig:Code Snippet}, showing how MiniApp collects and uses privacy. In the given MiniApp, \texttt{(a)(b)(c)} are part of the main package, which is initialized during a cold start, while \texttt{(d)(e)(f)} reside within the subpackage that is loaded on-demand when accessed. Since the subpackage code is only available during dynamic access, the static-only analysis may overlook privacy behaviors in \texttt{(d)}. Note that the privacy practices exhibited in \texttt{(b)(c)(d)} are potentially triggered by users, whereas the privacy behaviors in \texttt{(e)(f)} become dead code due to unused functions and orphaned pages, and thus will not be invoked. This implies that without UI state transition and callback control flow analysis, the privacy behaviors in \texttt{(e)(f)} might be incorrectly flagged as violations of the privacy policy, which is clearly a false positive. In the following, we describe this process in seven steps.

\noindent \ding{182} \textbf{Dynamic Data Binding~(JS $\Rightarrow$ WXML).} 
Dynamic data binding enables real-time updates of the user interface by linking JavaScript data directly to WXML widgets. In \texttt{pages/myInfo}, the \texttt{<navigator>} component specifies the navigation path as \texttt{\{\{takePhotoPath\}\}}, which is dynamically bound to the \texttt{data:takePhotoPath} in the logical layer.

\noindent \ding{183} \textbf{Main Package Internal Page Routing~(WXML $\Rightarrow$ WXML).} 
Page routing facilitates navigation between different views, enabling analysis to determine user-interactable pages.  In \texttt{pages/myInfo}, the \texttt{<navigator>} component directs the user to \texttt{data:takePhotoPath} when clicked, which is set to \texttt{``pages/takePhoto/takePhoto''} in the JavaScript logic.

\noindent \ding{184} \textbf{Dynamic-defined Event Handler Callback Control Flow~(WXML $\Rightarrow$ JS).} 
Event handler callback control flow is essential for identifying the program logic triggered by user actions. In \texttt{pages/takePhoto}, the \texttt{<button>} widget is bound to the event handler callback function \texttt{onShutterTap}, which is triggered when the button is tapped. Unlike statically defining the callback within the \texttt{Page} object, the callback function \texttt{onShutterTap} is dynamically defined by \texttt{util.init} in \texttt{onReady}.

\noindent \ding{185} \textbf{Cross-page Function Reuse Control Flow~(JS $\Rightarrow$ JS).} 
Cross-page function reuse is a critical aspect of maintaining code modularity and efficiency within a MiniApp. In \texttt{pages/takePhoto}, the event handler function \texttt{onShutterTap} is dynamically defined by the function \texttt{init} from \texttt{pages/util/util.js}, which is called with the current page context using \texttt{this}. This is achieved through context binding, ensuring that the \texttt{onShutterTap} has access to the correct page context when invoked.

\noindent \ding{186} \textbf{Data Transmission through Context Binding~(JS $\Rightarrow$ JS).} 
In \texttt{pages/takePhoto}, the function \texttt{takePhoto} invokes \texttt{util.onShutterTap} and passes the current page context (\texttt{this}) as an argument to the parameter \texttt{pageContext}. This binding allows the function \texttt{util.onShutterTap} to interact directly with the page's data and state. Within the function \texttt{onShutterTap}, \texttt{wx.createCameraContext} is invoked to initiate the photo-taking process. Upon successfully capturing an image, the callback \texttt{success} is triggered, and the resulting temporary file path (\texttt{res.tempImagePath}) is passed back to the page via \texttt{pageContext.setData}, setting the property \texttt{imagePath} to the file path.

\noindent \ding{187} \textbf{Main-to-Subpackage Page Routing~(JS $\Rightarrow$ WXML).} 
The navigation from \texttt{pages/myInfo} to \texttt{pages/checkID} is triggered by \texttt{navToCheckID}, which utilizes \texttt{wx.navigateTo} to specify the target path. Upon issuing the navigation command, the WeChat dynamically downloads the subpackage code and stores it in the local file system~(\texttt{com.tencent.mm/MicroMsg/hash/appbrand/pkg/} in Android). This dynamic loading ensures that the subpackage code is only fetched and loaded into memory when the user navigates to the corresponding page, thereby reducing the initial load time and memory usage of the MiniApp.

\noindent \ding{188} \textbf{Data Transmission through Event Propagation~(WXML $\Rightarrow$ JS).} 
Data transmission through event propagation is a fundamental concept in MiniApp development, facilitating the flow of user input from WXML to the underlying JavaScript logic. Here, the \texttt{bindinput} attribute of the \texttt{<input>} widget binds the \texttt{event:input} to the handler callback function \texttt{onInput}. The function \texttt{onInput} receives the \texttt{event} object as an argument, which contains the detailed user input value \texttt{event.detail.value}. This value is then used to construct a POST request to the server, with the parameter \texttt{idCard} set to the user input.

\subsection{Challenges for Hybrid Analysis of MiniApps}
\label{sec:challenges} 
Considering the unique nature of MiniApps, which relies on Web technologies and integrates with the capabilities of native apps, the hybrid analysis of MiniApps primarily faces three key challenges:

\noindent \textbf{Challenge\#1: Dynamic Data Binding across Different Languages in UI State Transition Analysis.} 
In MiniApps, dynamic data binding and page routing create complex interactions between JavaScript and WXML. This complexity arises from the need to accurately track how data updates~(target page path) propagate through the WXML components and JavaScript logic codes. Such an example is the step \ding{182} and \ding{183} in \cref{fig:Code Snippet}. Unlike Android where transitions are often explicitly defined through \texttt{Intent} objects and are relatively straightforward to trace, MiniApp involves a more intricate process due to the dynamic nature of data binding and cross-language interactions. Tracking and analyzing this type of UI state transition is challenging.

\noindent \textbf{Challenge\#2: Complexity of Cross-File Callback Control Flow Resolution.} 
Another prominent challenge emerges from the complexity of analyzing control flow, particularly due to the context binding when reusing callback functions across different files and subpackages. In MiniApps, callback functions often rely on the context~(\texttt{this}) in which they are executed. Such an example is the step \ding{184}, \ding{185} and \ding{186} in \cref{fig:Code Snippet}. Unlike traditional Android or iOS environments, where the scope and lifecycle of callbacks are more contained and predictable, MiniApps allow for the dynamic definition of callback functions, adding a layer of complexity. This approach makes it difficult to trace the origin and flow of these callbacks, as their definitions and bindings are spread across multiple files and only resolved dynamically.

\noindent \textbf{Challenge\#3: Inadequate in Locating WeChat-Specific Components in the UI Widgets Tree.}
To effectively load subpackages, we need to conduct a detailed UI exploration of the relevant pages. While this process may seem straightforward, the distinctive WXML components introduce unique challenges. These challenges primarily stem from the limitations of traditional testing tools. Designed specifically to align with the WeChat ecosystem, WXML components diverge significantly from the typical Document Object Model (DOM) structures or native UI frameworks familiar to conventional tools. This divergence renders tools like UIAutomator ineffective in identifying and locating these components directly from the GUI widgets tree. This limitation necessitates the development or adaptation of specialized testing tools and methodologies that can cater to the unique structure and behavior of WXML components, ensuring effective testing of MiniApps.

\subsection{Insights for Potential Solutions}
Upon evaluating the complexities inherent in MiniApps, particularly considering the aforementioned challenges, we distill three refined insights for potential solutions.

\noindent \textbf{Insight\#1: Integrating Cross-Language Data Flow to Constructing Complete UI State Transition Graph.}
To address Challenge\#1, the primary focus is on constructing a detailed and comprehensive topological structure of MiniApps,  with an emphasis on enhancing UI state transition analysis through the integration of cross-language data flows. When page routing is implemented through the \texttt{<navigator>} component or platform APIs, the target page paths need to be determined based on data flow analysis to identify the exact values. Especially when dynamic binding syntax is used to specify target paths in WXML, only a thorough data flow analysis of the JavaScript logic can accurately determine the destination page. This multifaceted approach captures the intricate interactions between WXML and JavaScript, ensuring a thorough understanding of data propagation and UI state transitions, thereby providing a robust foundation for page transition analysis and automated UI exploration.

\noindent \textbf{Insight\#2: Constructing Complete Cross-Page/Package Callback Control Flow by Analyzing Dynamic Definitions.}
To tackle Challenge\#2, the strategy involves integrating context binding directly into the callback control flow analysis. This approach focuses on analyzing how context~(\texttt{this}) propagates and is applied to define the callback functions across different files and packages.
By examining how callbacks are defined and where they are bound to specific contexts, we can create a more accurate and comprehensive representation of the callback control flow. 
Integrating context binding into the control flow provides a comprehensive understanding of the callback functions within a page, enabling better tracking of event-driven behaviors in MiniApps. 

\noindent \textbf{Insight\#3: Enhancing Existing Dynamic Testing Tools with Fuzzy Matching for Platform-Specific Components.}
In response to Challenge\#3, the strategy involves augmenting existing dynamic testing tools to better handle platform-specific MiniApp widgets. This enhancement would be achieved by implementing a fuzzy matching mechanism based on key component attributes like name, type, and text. By establishing an accurate correlation between static WXML components and dynamic UI Widget Tree elements, we can significantly improve the identification of MiniApp-specific widgets during dynamic runtime. This approach not only ensures the correct recognition of unique MiniApp components but also enhances the overall effectiveness of dynamic testing. 
\section{Methodology}
\label{sec:methodology}
\begin{figure*}[t]
    \centering   
    \includegraphics[width=0.95\linewidth]{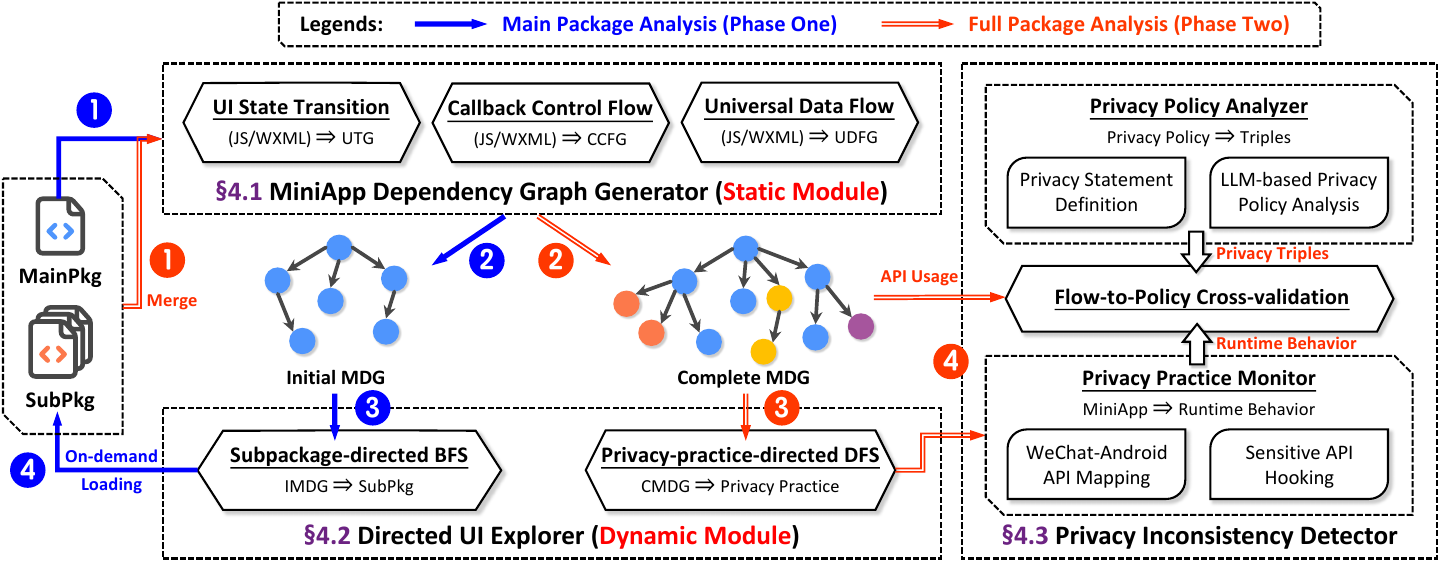}
    \caption{The architecture and workflow of \tool.}
    \label{fig:workflow}
\end{figure*}

In response to the three challenges, we propose \tool, an automated UI exploration and privacy inconsistency analysis framework based on two-phase iterative hybrid program analysis, as shown in \cref{fig:workflow}. 
The framework consists of three key modules: MiniApp Dependency Graph Generator~(\S\ref{sec:MDG Construction}), 
Directed UI Explorer~(\S\ref{sec:Runtime Behavior Verification}), and Privacy Inconsistency Detector~(\S\ref{sec:Privacy Compliance Detector}), which further includes a policy analyzer and a privacy practice monitor, thereby facilitating flow-to-policy cross-validation. Our methodology unfolds in two distinct phases, each involving a round of static analysis followed by a round of UI exploration. The entire workflow is structured as follows:

\noindent \textbf{Main Package Analysis (Phase One).} The initial phase of \tool is dedicated to dynamically loading subpackages through UI exploration to obtain the complete package. Firstly, \tool initializes the MiniApp Dependency Graph (MDG) for the main package. Subsequently, based on the guidance of MDG, the Directed UI Explorer employs a subpackage-directed breadth-first traversal strategy to explore subpackage pages, thus obtaining the complete package of the MiniApp.

\noindent \textbf{Complete Package Analysis (Phase Two).} In the second phase, \tool tackles the precise identification of privacy practices based on the analysis of the merged complete package. Mirroring the approach of the previous phase, we commence with constructing the complete MDG through static analysis. During the UI exploration stage, \tool adopts a privacy-practice-directed depth-first strategy, which is instrumental in conducting a runtime exploration of sensitive behaviors, allowing for a nuanced understanding of the MiniApp's privacy practices.

\subsection{MiniApp Dependency Graph Generator}
\label{sec:MDG Construction}
To facilitate automated UI exploration and more precise privacy practice detection, \tool analyzes the UI state transition, callback control flow and universal data flow. 
By combining them, \tool is able to derive a thorough and accurate topological view of a MiniApp.
\subsubsection{Analyzing UI State Transitions}
A MiniApp comprises numerous pages, each performing diverse functions. Interactions and switching between these modules are realized through page routing.
To precisely capture the page routing information, \tool constructs a UI State Transition Graph (UTG) to model this page routing logic. 

\noindent \textbf{Definition 1 (UTG):} UI State Transition Graph~(UTG) represents the static GUI model~\cite{yang2018wtg,dong2018frauddroid,liu2022promal}, which depicts page transition sequences in a MiniApp. Formally, the UTG can be expressed as a directed graph $G_U=(V_U, E_U, \lambda_U)$, where $V_U$ is a set of GUI states of pages with properties~(including type, resource ID, text, bounds, GUI event, and callback method); $E_U$ is a set of GUI events triggering page routing that act as the transition condition edges, and $\lambda_U$ is the function for parsing the page routing conditions.

\noindent\textbf{UTG construction.} During UTG construction, \tool analyzes the rendering and logic layers separately: (1) For the rendering layer, \tool focuses on the \texttt{<navigator>} widgets in the WXML file responsible for page navigation, with the \texttt{open-type} attribute signifying the routing method. (2) For the logic layer, \tool chiefly evaluates JavaScript code related to page routing APIs. For instance, \texttt{wx.navigateTo} adds the current page to the page stack and navigates to a new page. By examining five methods in the \texttt{<navigator>} widget and five types of page routing APIs in the logic layer as shown in \cref{tab:routing}, \tool can delineate all possible page transitions in the parsing function $\lambda_U$ and statically construct the UI state transition graph $G_U$. For statically defined page transition paths, \tool directly embeds these paths into the UTG. However, for dynamically bound page transitions, where the target page is represented as a variable rather than a hardcoded value, \tool records the variable name in the UTG as a placeholder. Subsequently, during data flow analysis, \tool resolves the value of such variables by tracking data propagation and value assignments in the code. This two-step approach ensures that dynamically defined transitions are accurately captured and added to the UTG once their target pages are resolved.

\begin{figure}[!htbp]
    \centering
    \includegraphics[width=0.75\textwidth]{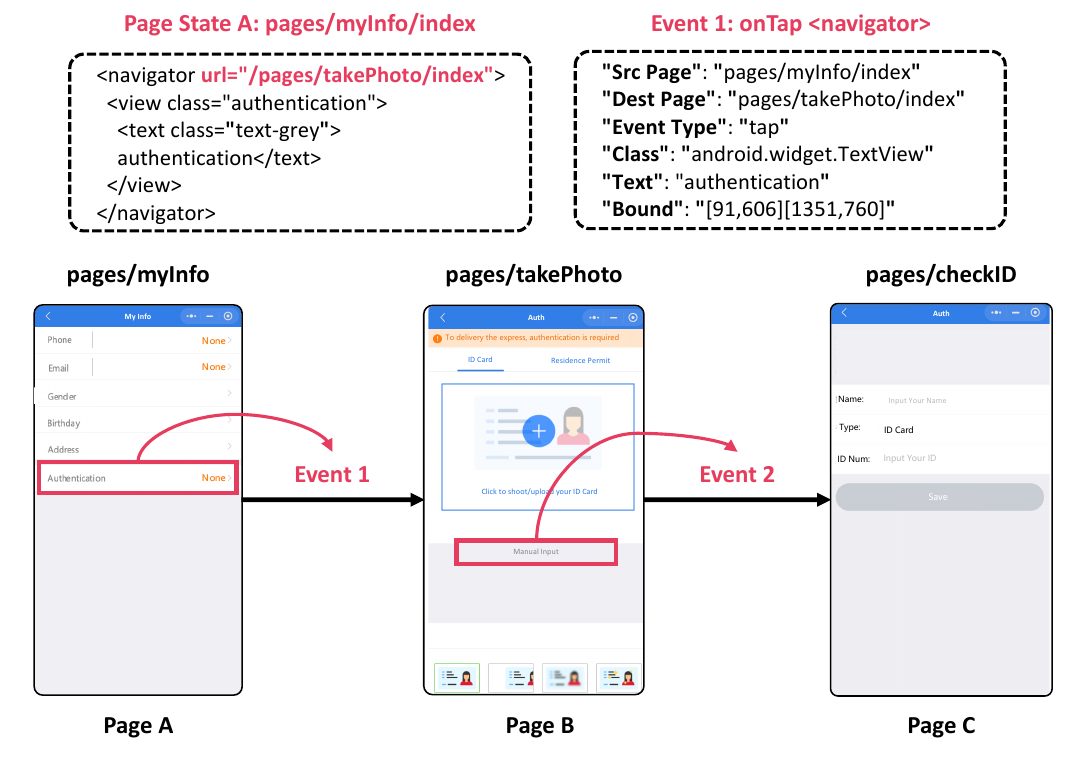}
    \caption{A simplified UTG example of \textit{Express} MiniApp. Corresponding to \cref{fig:Code Snippet}, in Event 1, the user triggers \texttt{ontap} event of \texttt{<navigator>}, leading to a transition from \texttt{pages/myInfo} to \texttt{pages/takePhoto}; in Event 2, the user triggers \texttt{ontap} event of \texttt{<button text=``manual input''>}, and the corresponding event handler callback function \texttt{naveToCheckID} is executed, \texttt{naveToCheckID} callback function, resulting in a transition from pages/myInfo to pages/checkID.}
    \label{fig:UTG Example}
\end{figure}

\noindent\textbf{A UTG example.} \cref{fig:UTG Example} illustrates a simplified UTG example of the \textit{Express} MiniApp. In this example, the set $V_U$ contains three pages: \texttt{``pages/myInfo/index''}(A), \texttt{``pages/takePhoto/index''}(B), and \texttt{``pages/checkID/index''}(C). In the MiniApp framework, all page widgets must be registered in the WXML file. We determine page states $V_U$ by traversing through these widgets. For example, \texttt{PageA} consists of six \texttt{<navigator>} widgets, and when we click on the \texttt{<navigator url=``/pages/takePhoto/index''>}, it will navigate to the corresponding page, making the widget a transition condition of the set $E_U$ from \texttt{PageA} to \texttt{PageB}. 

\begin{table}[!htbp]
    \centering
    \caption{MiniApp page routing methods considered in UI state transition resolution.}
    \fontsize{8}{12}\selectfont
    \begin{tabular}{ccc}
    \hline
    \textbf{Category} & \textbf{Property/Method} & \textbf{Description} \\
    \hline
    \multirow{6}{*}{<navigator>}    & navigate   & Keep the current page and navigate to a non-tabBar one. \\
    & navigateBack & Close the current page and go back. \\
    & redirect   & Close the current page and navigate to a non-tabBar one. \\
    & reLaunch   & Close all pages and open a new one. \\
    & switchTab  & Close all pages and open a target tabBar page. \\
    \hline
    \multirow{5}{*}{API}  & wx.navigateTo & Keep the current page and navigate to a non-tabBar one. \\
    & wx.navigateBack & Close the current page and go back to a previous one. \\
    & wx.redirectTo & Close the current page and navigate to a non-tabBar one. \\
    & wx.reLaunch   & Close all pages and open a new one. \\
    & wx.switchTab  & Close all pages and open a target tabBar page. \\             
    \hline
    \end{tabular}
    \label{tab:routing}
\end{table}

\subsubsection{Analyzing Callback Control Flows}
In MiniApps, rendering layer widgets bind to specific UI events. Upon occurrence of these events, the logic layer responds by executing the corresponding event handler callback function. Hence, we build the Callback Control Flow Graph~(CCFG) to conduct callback function level control flow analysis. 

\noindent \textbf{Definition 2 (CCFG):} Callback Control Flow Graph~(CCFG) models sequences of GUI event callbacks or lifecycle callbacks~\cite{yang2015ccfg}. Formally, CCFG can be expressed as a directed graph $G_C = (V_C, E_C, \lambda_C)$, where $V_C$ is a set of functions in the page of MiniApps, $E_C$ is a set of GUI events or lifecycle events that trigger callback functions that act as trigger condition edges, and $\lambda_C$ is the function for parsing the triggering conditions of callback functions.

\noindent \textbf{Determining CCFG entry point.} 
Prior to constructing, it is necessary to determine the CCFG entry points which are event-driven callback functions. 
This is accomplished by performing cross-language analysis on rendering layer widgets and their associated event handler callback functions. User interactions with these widgets trigger the corresponding event handling logic, activating the associated control flows within the MiniApp.
As an example shown in \cref{fig:Code Snippet}~(\ding{183}), a widget with \texttt{``bindtap = takePhoto''} attribute will bind to the \texttt{Tap} event, and execute the \texttt{takePhoto} function when a \texttt{Tap} event is triggered. Furthermore, the lifecycle callbacks can also serve as CCFG entry points as they are tied to specific lifecycle events, executing relevant logic when these events occur.

\noindent \textbf{CCFG construction.} 
From the identified entry points, \tool constructs the rest of the function call chain within the JavaScript language to obtain a complete CCFG.
Considering the highly customizable, modular nature of MiniApps, a tailored cross-file/package callback control flow analysis approach is necessary. 
Function definitions within MiniApp pages resemble class method definitions, as they're all defined within the \texttt{Page()} object instance. During initialization, developers provide all static-defined lifecycle callback functions, event handler functions, and custom functions to the \texttt{Page} object as attribute methods. Method intercommunication requires the \texttt{``this''} keyword, referring to the current \texttt{Page} object. 
Notably, some commonly used callback functions can be defined in shared JavaScript files and then dynamically imported across various pages.
Leveraging these observations, we design a cross-file callback control flow analysis algorithm based on the JavaScript Abstract Syntax Tree (AST). This algorithm takes into account the unique patterns of function usage and invocation in MiniApps, including the shared use of callback functions across different files and subpackages.

\noindent \textbf{Definition 3 (AST):} Abstract Syntax Tree (AST)~\cite{yamaguchi2014modeling} can represent the language structure of source code in detail. Formally, the AST can be expressed as an undirected property graph $G_A = (V_A, E_A, \lambda_A, \mu_A)$, where $V_A$ is a set of AST nodes, $E_A$ is a set of AST edges labeled by the function $\lambda_A$. 
Additionally, we assign a property to each node using $\mu_A$, which represents whether the node is an operator or an operand.

\noindent\textbf{Cross-File Callbacks Resolution Algorithm.} 
The following pseudocode in \cref{alg: CCFG} outlines the algorithm designed to identify and track cross-file callback definitions. The algorithm takes inputs including the AST $G_A$ of the current page and a set of $bindCalls$ obtained from the WXML. The output of the algorithm is the dynamically defined cross-file callback functions $exCalls$. 

\textbf{Step 1: Identify Imported Modules.} The algorithm first identifies all the modules imported into the current file, whether through \texttt{import} statements in ES6 or \texttt{require} calls in CommonJS. This is accomplished by the \texttt{getimportedModules} function. This function iterates over all nodes in the AST $G_A$. If a node is an \texttt{ImportDeclaration}, it records the module specified by the import statement~(Lines 3-5). If a node is a \texttt{VariableDeclarator} with an initialization that is a \texttt{CallExpression} calling \texttt{require}, it similarly records the module specified~(Lines 6-10). The function returns a dictionary \texttt{importedModules} mapping import specifiers to their sources~(Line 12).

\textbf{Step 2: Identify Cross-File Callbacks.} With the imported modules identified, the algorithm proceeds to identify cross-file callback functions. This is handled by the \texttt{getCrossfileCallbacks} function. This function iterates over the nodes in $G_A$ and looks for \texttt{CallExpression} nodes~(Line 16). If a \texttt{CallExpression} contains a \texttt{ThisExpression} in its arguments and the \texttt{callee} is one of the \texttt{importedModules} (Lines 16-17), it retrieves the position of the \texttt{ThisExpression} (Line 18) and the path of the module (Line 19). It then traverses the AST of the imported module to find matching \texttt{CallExpression} nodes and their contexts (Line 21). If it finds a \texttt{MemberExpression} with the correct context (Line 24) and the right child in \texttt{bindCalls} (Line 25), it appends the parent of this expression to \texttt{exCalls} (line 26). Finally, the function returns a node list of cross-file callback functions \texttt{exCalls}.

\begin{algorithm}[t]
\SetAlgoLined
\RestyleAlgo{ruled}
\caption{Cross-File Callbacks Resolution}
\label{alg: CCFG}
\fontsize{8}{10}\selectfont
\setlength{\itemsep}{0pt}
\setlength{\parskip}{0pt}
\setlength{\parsep}{0pt}

\KwIn {AST $G_A$ of the page, a set of $bindCalls$}
\KwOut {Dynamically defined cross-file callback functions $exCalls$}

% \SetKwFunction{FGetCCFG}{getCCFG}
\SetKwFunction{FgetimportedModules}{getimportedModules}
\SetKwFunction{FgetCrossfileCallbacks}{getCrossfileCallbacks}
\SetKwProg{Fn}{Function}{:}{end}

\Fn{\FgetimportedModules{$G_A$}}{
    \For{$node \in G_A.nodes $}{
        \If{$node$ is $ImportDeclaration$}{
            $ importedModules\left [node.specifier\right ] \gets node.source$  \\
        }
        \If{$node$ is $VariableDeclarator$}{
            \If{$node.init$ is $CallExpression$ \textbf{and} $node.init.callee.name$ is $require$}{ 
                $ importedModules[node.id.name] \gets node.init.arguments[0].value$ \\
            }
        }
    }
    \KwRet $importedModules$
}

\Fn{\FgetCrossfileCallbacks{$G_A, importedModules, pageMethods$}}{
    \For{$node \in G_A.nodes$}{
        \If{$node$ is $CallExpression$}{
            \If{$ThisExpression \in node.arguments$ \textbf{and} $node.callee \in importedModules.keys()$}{
            $pos \gets node.arguments.index(ThisExpression)$ \\
            $path \gets importedModules\left [node.callee\right ]$ \\
                \For{$exnode \in AST(path)$}{
                    \If{$exnode$ is $CallExpression$ \textbf{and} $exnode.callee==node.callee$}{
                        $ctx \gets exnode.argument.pos$ \\
                    }
                    \If{$exnode$ is $MemberExpression$ \textbf{and} $exnode.leftchild==ctx$}{
                        \If{$exnode.rightchild \in bindCalls$}{
                            $exCalls.append(exnode.parent)$ \\
                        }
                    }
                }
            }
        }
    }
    \KwRet $exCalls$
}

\end{algorithm}

\subsubsection{Merging into MinApp Dependency Graph}
To precisely model MiniApps, \tool combines UTG and CCFG with UDFG proposed in \textsc{TaintMini}~\cite{wang2023taintmini} to create a unified topological structure of MiniApps, called MiniApp Dependency Graph~(MDG). 
For example, \tool can analyze sensitive APIs distribution and their triggering paths by querying the graph.
In particular, we first formally describe UDFG following its original definition, then we define MDG accordingly. 

\noindent \textbf{Definition 4 (UDFG):} Universal Data Flow Graph (UDFG) can be used to capture the cross-language and cross-page data flow within a MiniApp~\cite{wang2023taintmini}. Formally, UDFG can be represented as a directed graph $G_D = (V_D, E_D, \lambda_D)$, where $V_D$ is the set of data objects, $E_D$ is the set of edges representing sensitive data flow dependencies and propagation directions. Correspondingly, $\lambda_D$ is the function for tracing data flow propagation.

\noindent \textbf{Definition 5 (MDG):} MiniApp Dependency Graph (MDG) is a joint topological structure combined by the above four directed graphs. The key insight for constructing MDG is that in each page's CCFG and UDFG, a node exists for each statement and predicate in the source code. It is natural to merge the page node of UTG with the corresponding root node of each page's AST. Therefore, MDG is formally represented as $G_M = (V_M, E_M, \lambda_M)$, where:

\begin{itemize}
    \item $V_M = \sum V_A \cup V_U$

    \item $E_M = \sum E_A \cup E_U \cup \sum E_C \cup \sum E_D$
    
    \item $\lambda_M = \lambda_A \cup \lambda_U \cup \lambda_C \cup \lambda_D$.
\end{itemize}

\begin{figure*}[!htbp]
    \centering
    \includegraphics[width=0.9\linewidth]{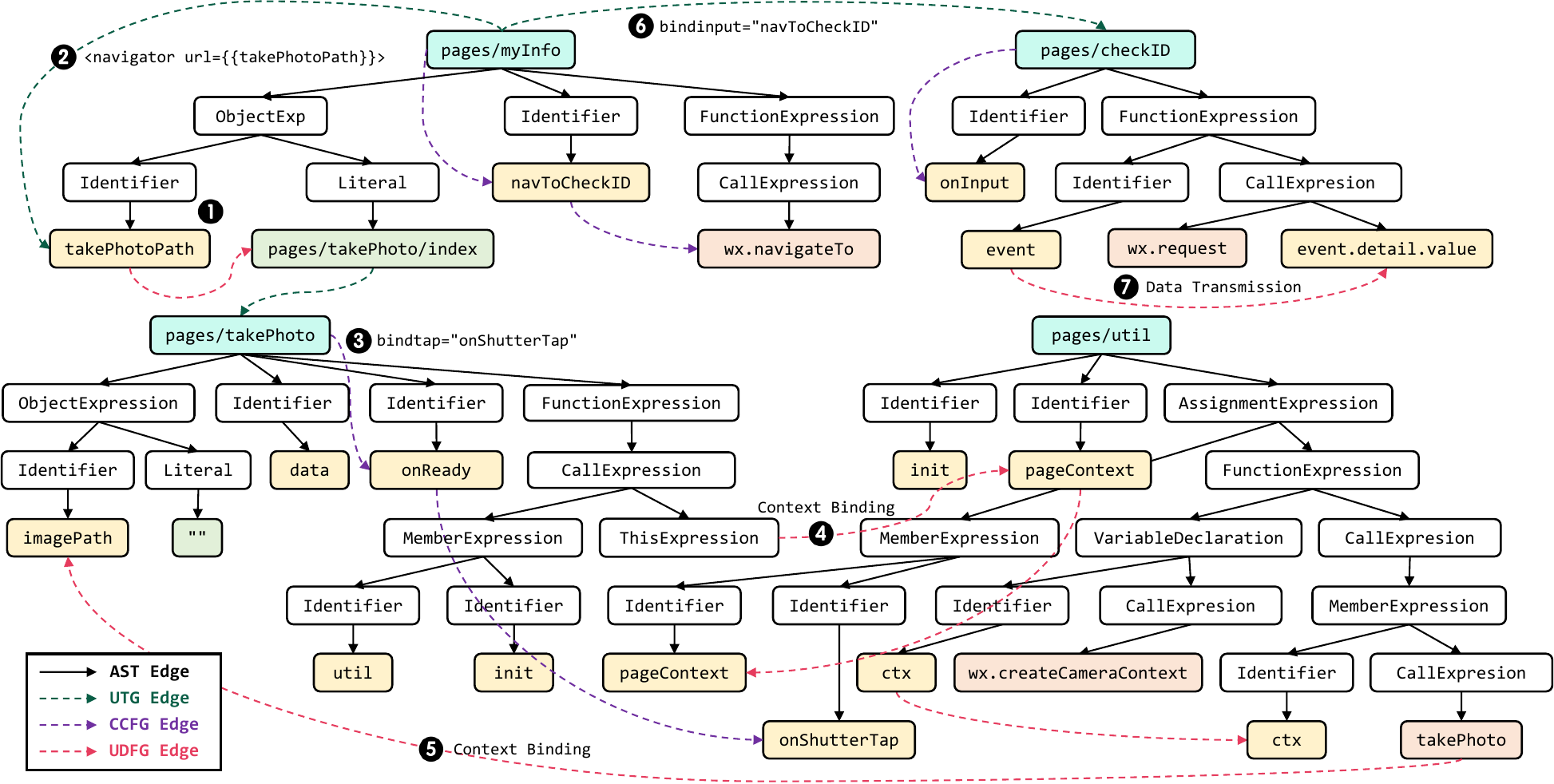}
    \caption{MDG representation of the code snippet presented in \cref{fig:Code Snippet}. The \textbf{Black} lines represent AST edges; the \textbf{\textcolor{UTG}{Green Lines}} represent UTG edges; the \textbf{\textcolor{CCFG}{Purple Lines}} represent CCFG edges; and the \textbf{\textcolor{UDFG}{{Pink Lines}}} represent UDFG edges. For simplicity, some AST nodes and edges have been omitted.}
    \label{fig:MDG Representation}
\end{figure*}

\noindent\textbf{MDG of the code snippet in the motivating example.} 
As shown in \cref{fig:MDG Representation}, the MDG captures multiple layers of interactions to analyze key behaviors, such as accessing camera resources. First, \tool identifies the navigation from \texttt{``pages/myInfo''} to \texttt{``pages/takePhoto''} via the \texttt{<navigator>} component, where the path is dynamically bound to the JavaScript variable \texttt{takePhotoPath}~(\ding{183} \textbf{\textcolor{UTG}{Green Line}}). This demonstrates dynamic data binding between the logical and rendering layers. 
Next, \tool identifies the event handler \texttt{onShutterTap}~(\ding{184} \textbf{\textcolor{CCFG}{Purple Line}}), which is bound to the \texttt{Tap} event of the \texttt{<button>} widget in \texttt{``pages/takePhoto''}. Unlike statically defined callbacks, \texttt{onShutterTap} is dynamically loaded by the \texttt{init} function in \texttt{pages/util/util.js}, which is invoked during the \texttt{onReady} lifecycle method. Through context binding~(\ding{185} \textbf{\textcolor{UDFG}{Pink Line}}), \tool can identify event handler callbacks defined across pages.
Through combined control and data flow analysis, \tool determines that the \texttt{onShutterTap} function invokes \texttt{wx.createCameraContext} to create a camera context, which is assigned to the variable \texttt{ctx}. Subsequently, the camera is triggered to take a photo using the \texttt{ctx.takePhoto} method. The captured image file path is then transmitted back to the page via \texttt{pageContext.setData}, dynamically updating the \texttt{imagePath} property.  

\subsection{Directed UI Explorer}
\label{sec:Runtime Behavior Verification}
Our dynamic analysis component leverages the constructed MDG to guide task-directed UI exploration. In the two-phase analysis, we respectively follow the subpackage-directed breadth-first exploration strategy and the privacy-practice-directed depth-first exploration strategy, thus achieving a balance between maximizing the UI state space and minimizing exploration time.

\subsubsection{Fuzzing Matching between WXML Component and UI Widget Tree}
As described in Challenge\#2, due to the platform-specific widgets defined in WXML (e.g., \texttt{<navigator>}) and the support for programmable attributes (e.g., loop rendering with \texttt{wx:for}), WXML components lack the unique \texttt{resourceID} and deterministic \texttt{XPATH} in the rendered UI screen, making it difficult to locate them uniquely. Therefore, to leverage the static MDG for guiding dynamic UI exploration, we propose a fuzzy matching approach to establish a robust mapping between the WXML component and the UI widget tree. 
The fuzzy matching comprises two primary strategies, as outlined below, to accurately locate widgets on the rendered UI screen. 

\begin{itemize}[leftmargin=2em]
    \item \textbf{Key Attribute Matching.} The first strategy involves the identification of a set of key attributes for each platform-specific WXML component, such as \texttt{name}, \texttt{type}, and \texttt{text}, etc. Then \tool employs the Intersection over Union (IoU) metric to match these key attributes with those of the widgets on the UI screen. The IoU is calculated as $IoU=\frac{W \cap U}{W \cup U}$, where $W$ represents a set of key attributes in the WXML component and $U$ represents a set of key attributes in the current UI secreen widgets. 

    \item \textbf{XPATH Matching.} In scenarios where UI widgets do not possess sufficient attribute information, or when the maximum IoU value between all widgets on the UI screen and the target WXML component falls below a 50\% threshold~\cite{yang2022permdroid}, \tool resorts to calculating the Levenshtein distance between the \texttt{XPATH} of target widget in WXML and widgets on the UI screen. The widget with the shortest Levenshtein distance is then selected as the candidate, followed by a verification of the UI state of the target page.
\end{itemize}

Through this fuzzy matching approach, \tool effectively bridges the gap between static MDG and dynamic UI exploration. By leveraging key attributes and the Levenshtein distance, \tool is able to accurately map WXML components to their corresponding widgets in the UI Widget Tree, overcoming the inherent challenges in locating platform-specific widgets.

\subsubsection{Task-directed UI Exploration Strategy}

\tool employs distinct exploration strategies in the two phases, meticulously designed to comprehensively analyze MiniApps by dynamically loading sub-packages and investigating privacy practices. 

\noindent \textbf{Subpackage-directed BFS Exploration.} 
In Phase One, to effectively load sub-packages and maximize the exploration of page state space, \tool utilizes the MDG of the main package for BFS exploration. This process begins with extracting all sub-package page transition paths from the MDG and maintaining them in a queue. Leveraging Appium~\cite{appium}, \tool obtains the current UI screen's widgets and calculates the IoU and Levenshtein distance to accurately locate the target widget within the transition path. Once navigation is executed, \tool verifies if the current page path aligns with the expected one. Upon reaching a target sub-package page, all related page transition paths are removed from the queue, streamlining the exploration process.
 
\noindent \textbf{Privacy-practice-directed DFS Exploration.} Once BFS is completed, \tool unpacks and merge the dynamically loaded sub-package into the main package. A subsequent round of static analysis updates the MDG. In Phase Two, guided by the complete MDG, \tool delves into a DFS exploration focusing on privacy-related practices within the MiniApp. 
Similar to BFS, this exploration process involves computing GUI events on specific pages to trigger relevant callbacks. However, in contrast to BFS, the DFS exploration prioritizes completing each potential privacy practice in a first-in, first-out order from the queue, ensuring a thorough examination.

Through the aforementioned two-phase task-directed UI exploration strategy, \tool achieves a comprehensive and systematic analysis of MiniApps, ensuring a robust and detailed understanding of MiniApp privacy practices, and addressing the challenges posed by their unique features.

\subsection{Privacy Inconsistency Detector}
\label{sec:Privacy Compliance Detector}

\tool could enable many security and software engineering tasks seamlessly. In this paper, our primary focus centers on the application of \tool to detect privacy inconsistency within MiniApps. Thus we monitor the privacy practices within MiniApps by instrumenting sensitive APIs and cross-validate with the statement in privacy policies to detect inconsistency.

\subsubsection{Privacy Practice Monitor}
In this component, \tool leverages the Frida framework~\cite{frida} to hook sensitive APIs. This enables us to monitor the API calls made by the MiniApp during dynamic UI exploration. When these API calls are triggered, our monitor captures crucial context information and stack traces, logging any detected statements. 
A key challenge arises from the architecture of MiniApps, where their logic layer primarily interacts with underlying functionalities through APIs encapsulated via the JSBridge mechanism~\cite{lu2020demystifying}. This encapsulation abstracts the connection between MiniApp APIs and the native system APIs, making it non-trivial to establish a direct mapping.
To address this, we drew insights from the prior work~\cite{lu2020demystifying} and replicated their approach as part of our analysis. Specifically, we started by reverse engineering the WeChat platform to analyze its JSBridge mechanism. We found that invoking a MiniApp API from JavaScript triggers a native bridge function offered by the host platform, which manages the execution of the actual system APIs. In the context of WeChat MiniApps, this bridge function is located within the \texttt{commonjni.AppBrandJSBridgeBinding} class, specifically in the invokeCallbackHandlerI(int, java.lang.String) method. Utilizing objection~\cite{objection}, a runtime mobile exploration toolkit powered by Frida~\cite{frida}, \tool hooks this bridge function, checking its parameters and runtime call stack. This allows us to establish the mapping from each sub-app API to the corresponding Java layer system API (detailed in our online documentation~\cite{online-form}). Note that this process of constructing the WeChat-to-Android API Mapping is offline and can be reused throughout our analysis once built.

\subsubsection{LLM-based Privacy Policy Analysis} 

To determine the disclosure of personal data collection and usage, we have drawn from the previous works of \textsc{Polisis}~\cite{harkous2018polisis} and \textsc{PolicyLint}~\cite{andow2019policylint}, then construct three ontologies to describe the privacy statement. 

\noindent \textbf{Definition of Privacy Statement.} As shown in \cref{tab:ontology}, we represent privacy policies as a series of triples, i.e., \texttt{(DC, SSoC, DE)}. \texttt{Data Controller} is the party responsible for determining the purposes and means of personal data processing, which can be the application itself (first party) or a third party. \texttt{SSoC Verbs} refer to a list of verbs that describe the \textbf{\underline{S}}toring-\textbf{\underline{S}}haring-\textbf{\underline{o}}r-\textbf{\underline{C}}ollection of data. \texttt{Data Entity} represents any privacy information or sensitive permission.
\begin{table}[!htbp]
    \centering
    \caption{Four types of ontologies in privacy policy.}
    \fontsize{8}{12}\selectfont
    \begin{tabular}{cc}
        \toprule
        \textbf{Labels} & \textbf{Ontology}\\
        \hline
        \textbf{DC} & Data Controller (First Party or Third Party) \\
        \textbf{SSoC} & Storing-Sharing-or-Collection Verbs\\
        \textbf{DE} & Data Entity (Privacy Information or Sensitive System Resources) \\
        \bottomrule
    \end{tabular}
    \label{tab:ontology}
\end{table}

\noindent \textbf{Few-shot Prompt Design.} 
Typically, previous works~\cite{harkous2018polisis,andow2019policylint,ling2022arethey} rely on manually annotated corpus to train named entity recognition models~(such as CRF, BiLSTM, BERT, etc.) for predicting entity labels, or by constructing data and entity dependency~(DED) trees to extract tuple representations of privacy policy statements. With the recent emergence of large language models~(LLMs), we aim to leverage the power of LLMs to extract fine-grained privacy statements from the policy. Instead of training the model from scratch, we utilize few-shot learning by providing the LLM with a small set of annotated examples~(few-shot demonstrations). These examples were carefully selected to cover a range of typical privacy policy statements involving different data controllers, actions, and data entities. Specifically, the prompt constructed consists of three parts: \textbf{(1) Task Description}, \textbf{(2) Few-shot Demonstrations}, and \textbf{(3) Query}. For better understanding, we provide the following prompt example. 

\begin{tcolorbox}[title=The Prompt Template, boxrule=0.8pt,boxsep=1.5pt,left=2pt,right=2pt,top=2pt,bottom=1pt]
\textbf{Task Description:} Here is a table where each row has 4 items. The first item is ``privacy statement'', and the second to fourth items are ``subject'', ``verb'', and ``object'' extracted from the ``privacy statement''. Here are some samples:

\textbf{Few-shot Demonstrations:} 

Demo\#1: ``To save photos, the developer will request your permission to access your photo album'', ``developer'', ``access'', ``photo album''.

Demo\#2: ``In order to help you become our member, the developer will collect your WeChat nickname and avatar after obtaining your express consent'', ``developer'', ``collect'', ``WeChat nickname and avatar''.

......

\textbf{Query:} Please read the table and follow the above sample rows to complete the table by filling in ``subject'', ``verb'', ``object'' of each row:
\texttt{\{Input Privacy Statement\}}.
\end{tcolorbox}

\subsubsection{Privacy Inconsistency Analysis}\label{subsec:consistence}\hfill

\noindent \textbf{API-to-Privacy Mapping.} To bridge the semantic gap between API calls and privacy policies, especially concerning data entities, we collect detailed descriptions of 29 sensitive APIs from the official WeChat documentation~\cite{Wechat_API_Documentation}, including API functions, parameters, etc. We correlate the API usage to specific privacy data access/collection based on semantics. That is, the data entities in privacy policies are divided into 13 categories, as detailed in our online documentation~\cite{online-form}.

\noindent \textbf{Flow-to-Policy Cross-validation.}
If sensitive data flows like collection, storage, or sharing are explicitly disclosed in the privacy policy, we consider them policy-compliant. Inconsistencies arise otherwise. Drawing from the consistency model in \textsc{PoliCheck}~\cite{andow2020actions}, we categorize inconsistent privacy disclosures into two: 1) Redundant Disclosures (RD), where the policy mentions more than actual privacy-related practices. This could mean developers disclose all possible sensitive behaviors, mitigating potential risks. 2) Omitted Disclosures (OD), where privacy-related practices occur but are not mentioned in the policy.
\section{Evaluation}
\label{sec:evaluation}

We have implemented a prototype of \tool using 9,466 lines of code (LoC) with Python, excluding any third-party libraries or open-source tools.
We then conduct an evaluation on \tool, aiming to answer the following research questions~(RQs):

$\bullet$ \textbf{RQ1 (Effectiveness):} How effective is \tool in detecting privacy-related practices and analyzing privacy policies? 

$\bullet$ \textbf{RQ2 (Ablation Study):} How does each component of \tool affect the performance separately? 

$\bullet$ \textbf{RQ3 (A Large-scale Study in the Real World):} How prevalent are privacy compliance violations in MiniApps ecosystem?

\subsection{Experimental Setup}

\noindent \textbf{Dataset.} To collect MiniApps, we utilize the open-source tool \texttt{MiniCrawler}~\cite{zhang2021measurement} to download MiniApp packages from the WeChat App Market. We have collected a total of 127,460 MiniApps, with 289 GB of total size. To collect privacy policies from our list of MiniApps, we design and deploy a privacy policy crawler. It requests privacy policies from the WeChat server based on the AppID, completing the collection process within 8 hours. Due to the prevalent absence of privacy policies, we find that only 10,786 (8.4\%) MiniApps have valid privacy policies.

\noindent \textbf{Ground Truth.} To ensure a reliable evaluation of RQ1 and RQ2, we carefully curate a comprehensive ground truth dataset by randomly sampling 100 MiniApps, which consists of two parts. To evaluate the performance of hybrid analysis, three experienced researchers separately interacted with these MiniApps for sub-package loading and manually inspected the privacy collected by the MiniApp, which forms the ground truth of their privacy-related practices. 
To ensure a thorough dynamic interaction, each researcher interacts with the MiniApp for 5 minutes or until all reachable pages/functions are fully triggered, whichever happens first. To evaluate the performance of LLM-based privacy policy analysis, we create a benchmark dataset by manually annotating 100 privacy policies associated with the sampled MiniApps. Specifically, three researchers independently read through each privacy policy and manually labeled the privacy practices mentioned in the text. This carefully curated benchmark, comprising 674 manually annotated privacy statements, serves as the ground truth for evaluating the performance of our privacy policy analysis techniques.

\noindent \textbf{Running Environment.} Our experiments are conducted on a server running Ubuntu Linux of 22.04 version with two 64-core AMD EPYC 7713 and 256 GB RAM. The static analysis leverages the server's computational capacity by utilizing 128 threads, enabling high parallelism for efficient processing of the large MiniApp dataset. The dynamic testing is performed on 16 Android Virtual Devices (AVDs) running in parallel, each configured with a system version of Android 8.1.0 and API Level 27. The version of WeChat used is 8.0.37, and the WebView kernel version is 107.0.5304.141.

%%%%%%%%%%%%%%%%%%%%%%%%%%%%%%%%%%%%%

\subsection{Effectiveness of \tool (RQ1)}
The efficacy of \tool, as outlined in \cref{sec:methodology}, is reliant on two key elements: 1) the accuracy of hybrid analysis; and 2) the effectiveness of privacy policy analysis. Our assessments of \tool focus on these two aspects.

\begin{table*}[t]
\def\arraystretch{1.1}
\centering
\caption{Effectiveness of hybrid analysis using \tool. GT represents the Ground Truth; Pre\% represents Precision\%; Rec\% represents Recall\%; F1\% represents F1-score\%.}
\label{tab:RQ1}
\small
% \begin{threeparttable}
{\fontsize{5.5}{8}\selectfont

\begin{tabular}{c|c|c||c|c|c|c|c|c||c|c|c|c|c|c}
\hline
\multirow{2}{*}{\textbf{Category}} & \multirow{2}{*}{\textbf{APIs}}   & \multirow{2}{*}{\textbf{GT}} & \multicolumn{6}{c||}{\textbf{\textsc{TaintMini}}} & \multicolumn{6}{c}{\textbf{\tool}} \\
\cline{4-15}
& & & \textbf{TP} & \textbf{FP} & \textbf{FN} & \textbf{Pre\%} & \textbf{Rec\%} & \textbf{F1\%} & \textbf{TP} & \textbf{FP} & \textbf{FN} & \textbf{Pre\%} & \textbf{Rec\%} & \textbf{F1\%} \\
\hline
\multirow{4}{*}{\textbf{Location}} & wx.chooseLocation                & 61                           & 60          & 16          & 1           & 78.9           & 98.4           & 87.6          & 61          & 0           & 0           & \textbf{100.0}          & \textbf{100.0}          & \textbf{100.0}         \\

& wx.getLocation                   & 168                          & 131         & 9           & 37          & 93.6           & 78.0           & 85.1          & 167         & 0           & 1           & \textbf{100.0}          & \textbf{99.4}           & \textbf{99.7}          \\

& wx.onLocationChange              & 22                           & 12          & 0           & 10          & \textbf{100.0}          & 54.5           & 70.6          & 17          & 1           & 4           & 94.4           & \textbf{81.0}           & \textbf{87.2}          \\

& wx.startLocationUpdateBackground & 9                            & 1           & 0           & 8           & \textbf{100.0}          & 11.1           & 20.0          & 4           & 1           & 4           & 80.0           & \textbf{50.0}           & \textbf{61.5}          \\
\hline                  

\multirow{4}{*}{\textbf{Media}}    & wx.chooseImage                   & 202                          & 178         & 21          & 24          & 89.4           & 88.1           & 88.8          & 202         & 1           & 0           & \textbf{99.5}           & \textbf{100.0}          & \textbf{99.8}          \\

& wx.chooseMedia                   & 13                           & 13          & 2           & 0           & \textbf{86.7}           & \textbf{100.0}          & \textbf{92.9}          & 13          & 3           & 0           & 81.3           & \textbf{100.0}          & 89.7          \\

& wx.chooseMessageFile             & 22                           & 20          & 1           & 2           & 95.2           & 90.9           & 93.0          & 21          & 0           & 1           & \textbf{100.0}          & \textbf{95.5}           & \textbf{97.7}          \\

& wx.chooseVideo                   & 23                           & 16          & 0           & 7           & \textbf{100.0}          & 69.6           & 82.1          & 20          & 0           & 3           & \textbf{100.0}          & \textbf{87.0}           & \textbf{93.0}          \\
\hline

\multirow{5}{*}{\textbf{OpenAPI}}  & wx.chooseAddress                 & 86                           & 76          & 2           & 10          & 97.4           & 88.4           & 92.7          & 84          & 0           & 2           & \textbf{100.0}          & \textbf{97.7}           & \textbf{98.8}          \\

& wx.chooseInvoiceTitle            & 7                            & 5           & 0           & 2           & \textbf{100.0}          & 71.4           & 83.3          & 6           & 0           & 1           & \textbf{100.0}          & \textbf{85.7}           & \textbf{92.3}          \\

& wx.getUserInfo                   & 60                           & 38          & 8           & 22          & 82.6           & 63.3           & 71.7          & 60          & 0           & 0           & \textbf{100.0}          & \textbf{100.0}          & \textbf{100.0}         \\

& wx.getUserProfile                & 171                          & 146         & 2           & 25          & 98.6           & 85.4           & 91.5          & 168         & 0           & 3           & \textbf{100.0}          & \textbf{98.2}           & \textbf{99.1}          \\

& wx.getWeRunData                  & 22                           & 13          & 0           & 9           & \textbf{100.0}          & 59.1           & 74.3          & 15          & 2           & 5           & 88.2           & \textbf{75.0}           & \textbf{81.1}          \\
\hline

\multirow{5}{*}{\textbf{Device}}   & wx.addPhoneContact               & 15                           & 2           & 0           & 13          & \textbf{100.0}          & 13.3           & 23.5          & 14          & 0           & 1           & \textbf{100.0}          & \textbf{93.3}           & \textbf{96.6}          \\

& wx.createCameraContext           & 17                           & 6           & 0           & 11          & \textbf{100.0}          & 35.3           & 52.2          & 15          & 0           & 2           & \textbf{100.0}          & \textbf{88.2}           & \textbf{93.8}          \\

& wx.createLivePusherContext       & 4   & 0           & 0           & 4           & 0.0            & 0.0            & 0.0           & 3           & 0           & 1           & \textbf{100.0}          & \textbf{75.0}           & \textbf{85.7}          \\

& wx.getRecordManager              & 6                            & 0           & 0           & 6           & 0.0            & 0.0            & /             & 6           & 0           & 0           & \textbf{100.0}          & \textbf{100.0}          & \textbf{100.0}         \\

& wx.openBluetoothAdapter          & 22                           & 1           & 0           & 21          & \textbf{100.0}          & 4.5            & 8.7           & 10          & 0           & 12          & \textbf{100.0}          & \textbf{45.5}           & \textbf{62.5}          \\
\hline

\multirow{2}{*}{\textbf{Album}}    & wx.saveImageToPhotosAlbum & 175  & 85          & 16          & 90          & 84.2           & 48.6           & 61.6          & 168         & 1           & 7           & \textbf{99.4}   & \textbf{96.0}           & \textbf{97.7} \\

& wx.saveVideoToPhotoAlbum  & 14  & 4  & 0 & 10          & \textbf{100.0}          & 28.6           & 44.4          & 12          & 0           & 2           & \textbf{100.0}          & \textbf{85.7}           & \textbf{92.3}          \\
\hline

\textbf{Total}                     & \textbf{/}                       & 1119                         & 807         & 77          & 312         & 91.3           & 72.1           & 80.6          & 1066        & 9           & 49          & \textbf{99.2}           & \textbf{95.6}           & \textbf{97.4}  \\
\hline
\end{tabular}
}

\end{table*}

\noindent \textbf{Effectiveness of Hybrid Analysis.}
We compare the performance of \tool with that of \textsc{TaintMini}~\cite{wang2023taintmini} on our ground truth dataset. We use the following metrics: True Positives (TPs), False Positives (FPs), False Negatives (FNs), precision, recall, and F1 score. 
The results of the evaluation are listed in \cref{tab:RQ1}. 
\tool significantly surpasses \textsc{TaintMini} in detecting 19 out of 20 sensitive APIs, offering the highest F1 score.
The lower precision of \textsc{TaintMini} stems from FPs found in unreachable code~(unused function and orphaned pages), while its decreased recall is due to its omission of sub-packages and its use of a single data flow analysis. 
For example, \textsc{TaintMini} only includes callbacks in its taint analysis when they are involved in data flows from WXML to JavaScript. This leads to \textsc{TaintMini} failing to determine the entry points of callback control flow in many cases, resulting in false positives related to unused functions. 
Furthermore, \textsc{TaintMini} overlooks many device-specific API calls like \texttt{wx.openBluetoothAdapter}, which may be called only once, not involving data flow propagation. \tool, by incorporating UTG, CCFG, and UDFG, provides the best overall performance.
However, it is important to note that \tool offers the lower precision compared to \textsc{TaintMini} for 4 specific APIs, which can be attributed to the path insensitivity of \tool's broader analysis scope. While subpackage loading provides a broader analysis scope, allowing \tool to detect more TPs, its path insensitivity leads to the merging of conditional branches, thereby introducing more FPs and affecting the precision metric. Although \tool effectively reduces most FPs caused by unused functions and orphaned pages, it cannot completely eliminate all of them due to its inherent path insensitivity.

\noindent \textbf{Time Efficiency of \tool.} In addition, we analyze the time efficiency of \tool to further evaluate its practicality. The analysis time is broken down into four components. First, the static analysis of the main package requires an average of 29.63 seconds per MiniApp, which involves analyzing the JavaScript and WXML code and extracting page transition, callback control flow, and data flow information. Second, sub-package-directed BFS UI exploration, which dynamically explores MiniApp pages based on the structure of sub-packages, takes approximately 64.97 seconds on average, depending on the number and depth of the sub-packages. Third, the complete package static analysis, which integrates the main package and sub-packages, requires around 36.54 seconds per MiniApp. Finally, privacy-practice-directed DFS UI exploration, which focuses on exploring privacy-relevant behavior and detecting sensitive API usage, is the most time-intensive phase, averaging 222.02 seconds per MiniApp. It is worth noting that \tool does not guarantee full UI coverage during testing, as covering all pages is unnecessary to achieve the testing objectives. These results indicate that the UI exploration phases, including sub-package-directed and privacy-practice-directed exploration, are the most time-consuming due to the complexity of dynamic exploration. Despite this, the time costs remain reasonable given the comprehensive analysis. 

\noindent \textbf{Effectiveness of LLM-based Privacy Policy Analysis.} We compare the performance of our LLM-based approach with five baseline NER models on the ground truth dataset of 100 manually annotated privacy policies. Specifically, we leverage GPT-3.5-\textsc{Turbo} to perform inquiries and extract privacy statements. During this process, we keep the default configuration of GPT-3.5-\textsc{Turbo}, with $temperature = 1$ and $top\_n =1$. To complete the experiment, we have utilized an estimation of 62k tokens in total to analyze 100 privacy policies. For the other five baseline models, we train them using annotations from the CA4P-483~\cite{zhao2022ca4p483} corpus. To ensure unbiased evaluation, we apply 10-fold cross-validation to the models, with 8 folds used for training, 1 fold for parameter tuning and optimization, and 1 fold for testing. We then leverage these baseline models to generate privacy statement annotations for the ground truth dataset. The performance of the LLM-based approach and the five baseline models is evaluated using standard metrics: Precision, Recall, and F1-score. The detailed performance results are presented in \cref{tab:rq1-baselines}. We observe that GPT-3.5-\textsc{Turbo} outperforms all five baseline models significantly, indicating its high robustness and effectiveness in extracting privacy practices from textual privacy policies.

\begin{table}[t]
\centering
\caption{Comparison with baseline NER models. DC represents Data Entity; SSoC represents Storing-Sharing-or-Collection Verbs; DE represents Data Entity. Pre\% represents Precision\%; Rec\% represents Recall\%; F1\% represents F1-score\%. }
{\fontsize{8}{12}\selectfont
\begin{tabular}{c||ccc||ccc||ccc}
\hline
\multirow{2}{*}{\textbf{Model}} & \multicolumn{3}{c||}{\textbf{DC}}  & \multicolumn{3}{c||}{\textbf{SSoC}}        & \multicolumn{3}{c}{\textbf{DE}}                           \\ \cline{2-10} 

& \multicolumn{1}{c|}{\textbf{Pre\%}} & \multicolumn{1}{c|}{\textbf{Rec\%}} & \textbf{F1\%} & \multicolumn{1}{c|}{\textbf{Pre\%}} & \multicolumn{1}{c|}{\textbf{Rec\%}} & \textbf{F1\%} & \multicolumn{1}{c|}{\textbf{Pre\%}} & \multicolumn{1}{c|}{\textbf{Rec\%}} & \textbf{F1\%} \\ \hline

\textbf{HMM}                    & \multicolumn{1}{c|}{21.4}           & \multicolumn{1}{c|}{73.9}           & 33.2          & \multicolumn{1}{c|}{27.9}           & \multicolumn{1}{c|}{82.9}           & 40.4          & \multicolumn{1}{c|}{43.0}           & \multicolumn{1}{c|}{74.6}           & 54.6          \\ \hline
\textbf{CRF}                    & \multicolumn{1}{c|}{64.2}           & \multicolumn{1}{c|}{52.9}           & 58.0          & \multicolumn{1}{c|}{72.9}           & \multicolumn{1}{c|}{72.0}           & 72.4          & \multicolumn{1}{c|}{74.2}           & \multicolumn{1}{c|}{77.1}           & 75.6          \\ \hline
\textbf{BiLSTM}                 & \multicolumn{1}{c|}{61.0}           & \multicolumn{1}{c|}{65.3}           & 63.1          & \multicolumn{1}{c|}{78.6}           & \multicolumn{1}{c|}{79.4}           & 78.9          & \multicolumn{1}{c|}{80.4}           & \multicolumn{1}{c|}{75.0}           & 77.6          \\ \hline
\textbf{BiLSTM-CRF}             & \multicolumn{1}{c|}{66.9}           & \multicolumn{1}{c|}{68.7}           & 67.8          & \multicolumn{1}{c|}{79.0}           & \multicolumn{1}{c|}{80.0}           & 79.4          & \multicolumn{1}{c|}{82.0}           & \multicolumn{1}{c|}{76.8}           & 79.3          \\ \hline
\textbf{BERT}                   & \multicolumn{1}{c|}{65.5}           & \multicolumn{1}{c|}{68.1}           & 66.3          & \multicolumn{1}{c|}{77.1}           & \multicolumn{1}{c|}{85.6}           & 81.1          & \multicolumn{1}{c|}{79.3}           & \multicolumn{1}{c|}{82.1}           & 80.7          \\ \hline

\textbf{GPT-3.5-\textsc{Turbo}}                 & \multicolumn{1}{c|}{\textbf{100}}          & \multicolumn{1}{c|}{\textbf{100}}          & \textbf{100}         & \multicolumn{1}{c|}{\textbf{95.4}}           & \multicolumn{1}{c|}{\textbf{91.4}}           & \textbf{93.4}          & \multicolumn{1}{c|}{\textbf{100}}          & \multicolumn{1}{c|}{\textbf{91.8}}           & \textbf{95.7}          \\ \hline
\end{tabular}
}
\label{tab:rq1-baselines}
\end{table}

\begin{tcolorbox}[title=ANSWER to RQ1, boxrule=0.8pt,boxsep=1.5pt,left=2pt,right=2pt,top=2pt,bottom=1pt]

For hybrid analysis, \tool outperforms \textsc{TaintMini} on our ground truth dataset with 99.2\% precision, 95.6\% recall, and 97.4\% F1 score. Similarly, in privacy policy analysis, our LLM-based method surpasses existing NER models with 98.5\% precision, 94.5\% recall, and 96.5\% F1 score.

\end{tcolorbox} 

\subsection{Ablation Study (RQ2)}
We conduct an ablation study to understand how UTG and CCFG improve \tool's performance. Three \tool variants are developed for this: 1) \tool-UDFG-ONLY, which disables UTG and CCFG to emulate the \textsc{TaintMini} approach; 2) \tool-STATIC-ONLY, which enables UTG and CCFG; and 3) \tool-DYNAMIC-ONLY, which removes static guidance. 

\begin{figure}[t]
    \centering
    \includegraphics[width=0.35\linewidth]{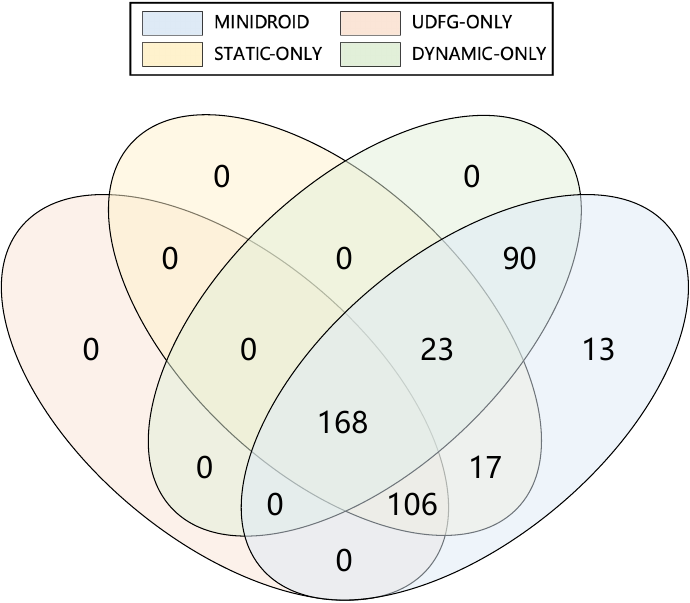}
    \caption{Venn diagram showing the privacy-related practices detected by each component of \tool.}
    \label{fig:rq2-venn}
\end{figure}

\noindent\textbf{Number of Privacy-related Practice Detected by \tool Variants.}
\cref{fig:rq2-venn} illustrates the number of privacy-related practices detected by \tool and three variant baselines on the ground truth dataset. It can be observed that \tool is a superset of each variant baseline. Furthermore, by combining static and dynamic approaches, \tool identifies 13 privacy-related practices that cannot be detected by each individual component. This indicates that the methodology adopted by \tool improves the performance of privacy-related practice identification and also demonstrates its robustness.

\begin{figure}[t]
    \centering
    \includegraphics[width=0.65\linewidth]{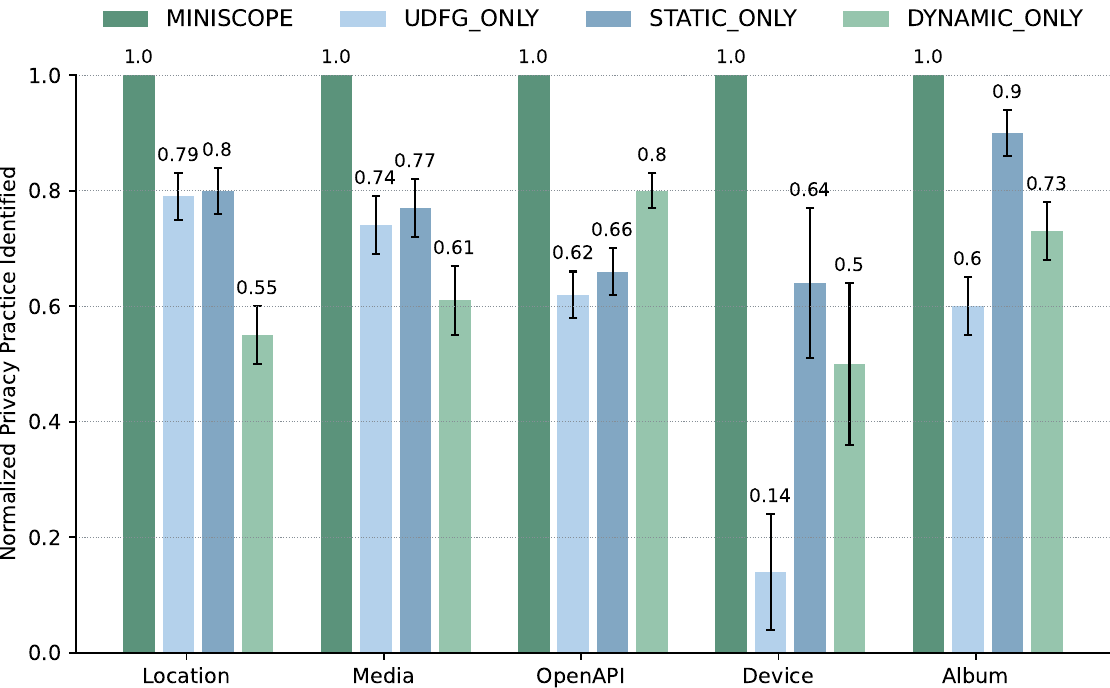}
    \caption{The performance of \tool, \tool-UDFG-ONLY, \tool-STATIC-ONLY and \tool-DYNAMIC-ONLY on normalized privacy-related practice identified.}
    \label{fig:ablation study}
\end{figure}

\noindent\textbf{Normalized Privacy-related Practice Detection by \tool Variants.}
For better visualization and understanding, we present the normalized privacy-related practices detected by \tool and other three variant baselines in \cref{fig:ablation study}. The normalized value is computed as follows: $nomalized\_privacy\_practice = \frac{pp}{pp(\tool)}$. 

From the comparison between UDFG-ONLY and STATIC-ONLY, we notice a significant performance boost in static privacy-related practice identification due to the integration of UTG and CCFG, especially in the \texttt{Device} and \texttt{Album} categories. This reinforces our RQ1 findings where \textsc{TaintMini}'s dependence on single data flow analysis leads to increased FNs in these API categories. STATIC-ONLY generally outperforms DYNAMIC-ONLY. This is a result influenced by MiniApps' characteristics, such as the need for user login or specific purchase completions, which impact dynamic analysis recall. By utilizing hybrid analysis, \tool efficiently combines their strengths for more effective privacy practice identification.

\begin{tcolorbox}[title=ANSWER to RQ2, boxrule=0.8pt,boxsep=1.5pt,left=2pt,right=2pt,top=2pt,bottom=1pt]
By conducting an ablation study on each component, we discover that the hybrid analysis employed by \tool indeed enhances the performance of privacy-related practice identification while also demonstrating robustness.
\end{tcolorbox} 

\subsection{A Large-scale Study in the Real World (RQ3)}

\begin{table}[ht]
\centering
\caption{Privacy inconsistencies detected by \tool.}
{\fontsize{8}{12}\selectfont
\begin{tabular}{c|c|c||cc||cc}
\hline
\multirow{2}{*}{\textbf{Category}} & \multirow{2}{*}{\textbf{Scopes}} & \multirow{2}{*}{\textbf{Total}} & \multicolumn{2}{c||}{\textbf{Redundant}}            & \multicolumn{2}{c}{\textbf{Omitted}}              \\ \cline{4-7} 

&   &     & \multicolumn{1}{c|}{\textbf{Count}} & \textbf{Percent\%} & \multicolumn{1}{c|}{\textbf{Count}} & \textbf{Percent\%} \\ 
\hline

\textbf{Location}                  & UserLocation                     & 7,714                            & \multicolumn{1}{c|}{1,389}           & 18.1        & \multicolumn{1}{c|}{64}             & 0.9         \\ \hline

\textbf{Media}                     & ChooseMedia/File                 & 13,115                           & \multicolumn{1}{c|}{1,594}           & 12.2        & \multicolumn{1}{c|}{193}            & 1.5         \\ \hline
\multirow{4}{*}{\textbf{OpenAPI}}  & Address                          & 8,714                            & \multicolumn{1}{c|}{1,063}           & 12.2        & \multicolumn{1}{c|}{126}            & 1.5         \\ 

& Invoice                          & 3,540                            & \multicolumn{1}{c|}{490}            & 13.9        & \multicolumn{1}{c|}{7}              & 0.2         \\ 
& UserInfo                         & 14,535                           & \multicolumn{1}{c|}{1,380}           & 9.5         & \multicolumn{1}{c|}{116}            & 0.8         \\ 
& WeRun                            & 2,297                            & \multicolumn{1}{c|}{489}            & 21.3        & \multicolumn{1}{c|}{8}              & 0.4         \\ 
\hline

\multirow{6}{*}{\textbf{Device}}   & PhoneContact                     & 1,179                            & \multicolumn{1}{c|}{163}            & 13.9        & \multicolumn{1}{c|}{24}             & 2.1         \\ 

& PhoneCalendar                    & 796                             & \multicolumn{1}{c|}{395}            & 49.7        & \multicolumn{1}{c|}{/}              & /            \\

& Camera                           & 8,283                            & \multicolumn{1}{c|}{809}            & 9.8         & \multicolumn{1}{c|}{9}              & 0.2         \\

& Record                           & 4,773                            & \multicolumn{1}{c|}{849}            & 17.8        & \multicolumn{1}{c|}{1}              & 0.1         \\

& Bluetooth                        & 1,941                            & \multicolumn{1}{c|}{494}            & 25.5        & \multicolumn{1}{c|}{13}             & 0.7         \\

& Clipboard                        & 54                              & \multicolumn{1}{c|}{15}             & 27.8        & \multicolumn{1}{c|}{/}              & /            \\ 
\hline

\textbf{Album}                     & PhotoAlbum                       & 11,198                           & \multicolumn{1}{c|}{1,095}           & 9.8         & \multicolumn{1}{c|}{250}            & 2.3         \\ \hline

\textbf{Total}                     & \textbf{/}                       & 78,139                           & \multicolumn{1}{c|}{10,225}          & 13.1        & \multicolumn{1}{c|}{811}            & 1.1         \\ \hline
\end{tabular}
}
\label{tab:rq3}
\end{table}

\noindent \textbf{Measurement.} We utilize \tool to conduct a large-scale compliance violation detection on 127,460 MiniApps in the real world. The preliminary results indicate that out of these MiniApps, 33,570 (26.37\%) collect and use privacy information. However, only 10,786 (8.46\%) of these MiniApps have valid privacy policies in place. This highlights the severe privacy risks prevalent in the MiniApp ecosystem. Furthermore, we assess the consistency of privacy-related practices against privacy policies to detect privacy compliance violations. 
\cref{tab:rq3} presents the detailed statistical results of the compliance detection. Redundant Disclosure refers to disclosures in a privacy policy that exceed privacy-related practices, while Omitted Disclosure refers to disclosures in a privacy policy that are fewer than privacy-related practices. The former is considered a weak violation, while the latter is considered a strong violation. Due to the possibility of multiple instances of RD or OD in a MiniApp, we classify and record the distribution of disclosures for each type of privacy-related practice. Overall, we summarize our findings as follows.

\noindent \textbf{Findings.} 
In our study of compliance violations in MiniApps, we find that Redundant Disclosure (13.1\% of all cases) significantly outweighs Omitted Disclosure (1.1\%). MiniApps tend to excessively disclose privacy-related practices linked to \texttt{PhoneCalendar} (49.7\%) more often, possibly due to developers retaining policy disclosures even after removing associated privacy practices once time-limited events conclude. In contrast, Omitted Disclosure is more prevalent in categories like \texttt{Media}, \texttt{Address}, \texttt{UserInfo}, and \texttt{PhotoAlbum}, likely due to developers not realizing these categories require permission as per WeChat's policy~\cite{Wechat_API_Documentation}. Our large-scale study reveals that 5.7\% (614/10786) of MiniApps over-collect data secretly, while 33.4\% (3599/10786) overstate their actual data collection.

\noindent \textbf{Responsible Disclosure to Developers.} Based on our findings, we have responsibly disclosed these violations to their developers via the email obtained from the privacy policies. Particularly, we shared our methodology and the trigger paths of potential privacy compliance violations detected by \tool with developers and asked for their feedback on these findings. Overall, we notified a total of 2,282 developers, out of which 1,727 emails are successfully sent, 396 emails are intercepted by the server’s filtering policies, and 159 of them are reported as undeliverable email addresses. 
As shown in \cref{tab:rq3-response}, as of the time of writing, we have received 46 responses from MiniApp developers. 
We summarize their responses as follows:

\begin{table}[!t]
\centering
\caption{Responses from developers.}
\fontsize{8}{12}\selectfont
\begin{tabular}{c|c|c|c}
\hline
\textbf{Response}  & \textbf{Measures Taken} & \textbf{Count by Each Type} & \textbf{Total} \\ \hline
\multirow{2}{*}{\textbf{Acknowledgement}} & accept and update the privacy policy     &     overclaim (25) \& over-collection (17)     & 42             \\  
\cline{2-4}
                                          & partially accept and update the privacy policy &  overclaim (2)  & 2            \\ 
\hline
\textbf{Disagreement}                     & claim their reasons   &  overclaim (2)      & 2 \\ \hline
\end{tabular}
\label{tab:rq3-response}
\end{table}

\begin{itemize}[leftmargin=1.2em]
    \item \textit{42 developers fully accept our findings and commit to updating their privacy policies.} Most acknowledge their privacy policies are outdated, with redundant disclosures stemming from past versions that incorporated relevant privacy practices~(which further confirms our previous findings). Developers express a desire for the MiniApp platform to offer automated code audits and maintain consistent privacy compliance.
    
    \item  \textit{2 developers partially accept our findings and provide valuable suggestions.} They attribute some redundant disclosures to user-input data collection via forms in MiniApps. Thus, they consider our findings partially accurate, intending to selectively update based on our results. These insights underscore potential areas of improvement in our approach, particularly in dealing with user-input privacy within MiniApps. 
    
    \item  \textit{2 developers disagree with our findings.} These developers highlight that a portion of the detected privacy over-claims in our study stem from selective functionality accessibility, which is specifically tailored for certain target users or for time-limited promotions. This selective accessibility can be implemented based on special entry points, such as QR code scanning, leading to the scenario that our analysis tool is unable to access restricted pages within the MiniApps, thereby omitting the analysis of corresponding privacy practices. We further discuss these cases in \S\ref{sec:threats-to-validity}.
\end{itemize}

\noindent \textbf{Case Studies.} To understand the impacts of privacy violations, we have conducted case studies by inspecting these inconsistent MiniApps. In the following, we present two typical cases:

\begin{itemize}[leftmargin=1.2em]
    \item \textbf{Redundant Disclosure.} An example of extreme redundant disclosure is an online ordering MiniApp named \texttt{WanMoTang(AppID: wx5e4dc66b2f***)}, with a 4.2-star rating and over 1k recent users. Although its privacy policy discloses nearly all privacy types (19 in total), the MiniApp only employs 3 types in practice: UserInfo, Location, and Photoalbum. This over-disclosure potentially escalates privacy risks, as any privacy practice may seem justifiable against an overly comprehensive privacy policy.
    
    \item  \textbf{Omitted Disclosure.} 
    One instance of omitted disclosure in the MiniApp \texttt{HaiHuiShou(AppID: wxded0379a82***)}, which has an average 4.0 rating and around 200 reviews. Although it discloses 4 privacy types in its policy, our analysis uncovered undisclosed practices in its \texttt{pages/demo/demo} source code. The \texttt{onLoad} function creates a \texttt{Camera} context, requests \texttt{WeRun} data, and collects \texttt{Address} information. These undisclosed practices constitute privacy compliance violations.
\end{itemize}

\begin{tcolorbox}[title=ANSWER to RQ3, boxrule=0.8pt,boxsep=1.5pt,left=2pt,right=2pt,top=2pt,bottom=1pt]
In our extensive evaluation of 127,460 MiniApps, we found that over 91.5\% lack effective privacy policies, and of those that do, 39.1\% exhibit flow-to-policy inconsistencies. After responsibly disclosing these findings to 2,282 developers, we received confirmations and acknowledgments from 44 of them.

\end{tcolorbox} 
\section{Discussion}

\subsection{Threats to Validity}
~\label{sec:threats-to-validity}

\noindent\textbf{Influence from unpacking of MiniApps.} Despite our attempts to utilize state-of-the-art open-source tools such as wxappUnpacker~\cite{wxappUnpacker} and unveilr~\cite{unveilr}, 235 (0.2\%) of the MiniApps in our dataset cannot be successfully unpacked. 
Certain MiniApps, developed using third-party multi-end frameworks for a uniform programming style, may not have a standard page structure after unpacking. This situation, affecting around 1.3\% of our ground truth dataset, contributes to an approximately 11.0\% false negative rate in static-only analysis.

\noindent\textbf{Corner cases encountered in the dynamic analysis.} The effectiveness of the dynamic analysis could be hampered by unexpected corner cases. 
For example, in online shopping MiniApps, where completing orders or accessing review pages requires user interaction or external credentials, we rely on static-only analysis to detect potential privacy inconsistencies. Similarly, certain pages or functions are dynamically made available to users based on specific criteria, such as user permissions or account status. Since these pages are not accessible during dynamic testing, the analysis may miss some functionality or behavior that depends on runtime conditions.

\noindent \textbf{Potential misclassification of orphaned pages.} During the disclosure process, two developers raised concerns about the classification of certain pages as orphaned. Specifically, they pointed out that some pages, while not referenced in the navigation structure, are intentionally made accessible through non-standard entry points, such as QR code scanning or shared links. This limitation in static analysis could lead to potential misclassification of these pages and their associated functions, resulting in false positives when identifying privacy-policy inconsistencies. For example, if a page is incorrectly marked as orphaned and its related code is treated as dead, \tool may miss actual privacy behaviors associated with these pages, leading to potential false positives in redundant privacy disclosures. Due to current technical limitations, our approach cannot reliably distinguish genuinely orphaned pages from those accessible via non-standard entry points, which remains an open challenge for future research.

\noindent\textbf{Possible bias in the ground truth.} The ground truth used in our evaluation (RQ1 and RQ2) may be subject to manual confirmation bias. Despite our efforts to involve three experienced security researchers to conduct meticulous inspections and reach a consensus, there may still be potential biases, which could arise in sub-package dynamic loading or privacy-related practices identification.

\subsection{Scalability and Transferability}
In our research, while the primary focus is on the WeChat MiniApp ecosystem similar to other works~\cite{wang2023taintmini,meng2023wemint,li2023minitracker}, it is important to emphasize the scalability and transferability of our proposed methodology to other platforms.
As highlighted in the W3C MiniApp Standardization White Paper~\cite{w3c}, MiniApps across most platforms share similar underlying architectures, employing JavaScript for logic programming and analogous layout files (WXML in WeChat, AXML in Alipay, SWAN in Baidu, and TTML in TikTok).
This architectural uniformity suggests that \tool, initially tailored for WeChat, holds significant potential for broader applicability with minimal adjustments. 
% To extend our framework to these different MiniApp ecosystems, it would primarily requires adapting to the specific components and APIs unique to each platform's MiniApp framework.
By fine-tuning it to accommodate the nuances of each platform's specific components and APIs, \tool can be effortlessly migrated and applied to other ecosystems beyond WeChat. 
This transferability not only enhances the utility and reach of our research but also opens avenues for comprehensive privacy and security analysis across diverse MiniApp platforms.

\section{Related Work}
MiniApps, as a novel application paradigm, have begun to attract scholarly attention in recent times. Prior studies in this field can be categorized into three primary aspects.

\noindent \textbf{Security and Privacy of MiniApps.}
Several investigations have delved into the security aspects of Miniapps~\cite{zhao2023signature,wang2023uncovering,zhang2023small,liu2020industry,zhang2024minicat}. Wang et al.~\cite{wang2022characterizing} gathered 83 MiniApp bugs from real-world scenarios and developed \textsc{WeDetector} to identify WeBugs following three bug patterns. Another work~\cite{lu2020demystifying} probed into issues like system resource exposure, subwindow deception, and sub-app lifecycle hijacking within the Mini-Program ecosystem. They conducted evaluations on 11 popular platforms to ascertain the widespread nature of these security problems. Besides, Zhang et al.~\cite{zhang2022identity} systematically studied the identity confusion vulnerability in WebView-based app-in-app ecosystems, revealing how improper identity checks could allow MiniApps to misuse privileged APIs, leading to potential privacy breaches. Yang et al.~\cite{yang2022cross} identified the Cross Mini-program Request Forgery (CMRF) vulnerability caused by missing AppID checks in MiniApp communication, and proposed CmrfScanner, a static analysis tool, to detect this issue at scale.  
%%%%%%%%%%%%%%%%%%%%
Additionally, a series of studies have emphasized the importance of privacy in MiniApp ecosystem~\cite{yang2023sok,zhang2023spochecker,wang2023doasyousay,li2023minitracker,wang2023sats,wang2024minichecker,yan2023muid}. \textsc{TaintMini}~\cite{wang2023taintmini} introduced a framework for detecting flows of sensitive data within and across mini-programs using static taint analysis. Another work MiniTracker~\cite{li2023minitracker} constructed assignment flow graphs as common representation across different host apps and performed a large-scale study on 150k MiniApps, which revealed the common privacy leakage patterns. Moreover, several studies~\cite{meng2023wemint,baskaran2023measuring,zhang2023dont} have focused on taint analysis to detect AppSecret leaks. In particular, another work~\cite{wang2023doasyousay} focused on the consistency of data collection and usage in MiniApps. They crawled 2,998 MiniApps and detected 89.4\% of them violated their privacy policies. More recently, Zhang et al.~\cite{zhang2023spochecker} introduced SPOChecker and performed the first systematic study of privacy over-collection in MiniApps. 
As listed in \cref{tab:related work}, despite these significant contributions to understanding MiniApp privacy dimensions, these approaches fall short in accurate privacy behavior identification.

\begin{table}[!htbp]
\fontsize{8.5}{12}\selectfont
\caption{Comparison of \tool with other tools. Note that \Circle ~indicates that the tool does not support this feature; \RIGHTcircle ~signifies that the tool supports this feature, but there is a gap when compared to the SOTA; \CIRCLE ~denotes the SOTA available.}
\begin{threeparttable}
\begin{tabular}{ccccccc}
\toprule
\multirow{2}{*}{Approach} & \multicolumn{3}{c}{Static Analysis}                              & \multicolumn{3}{c}{Dynamic Analysis}          \\
                          & UI Transition & CFA & DFA \& Taint & Subpackage & UI Exploration & Instrumentation \\
\hline
\textsc{TaintMini}~\cite{wang2023taintmini}  &  \Circle  &  \RIGHTcircle  & \CIRCLE  &  \Circle  &  \Circle  &  \Circle  \\
\textsc{MiniTracker}~\cite{li2023minitracker}  &  \Circle  &  \RIGHTcircle  &  \CIRCLE  &  \Circle  &  \Circle  &  \Circle \\
\textsc{WeMinT}~\cite{meng2023wemint}  &  \Circle  &  \Circle  &  \CIRCLE  &  \Circle  &  \Circle  &  \Circle \\
\textsc{SPOChecker}~\cite{zhang2023spochecker}  &  \RIGHTcircle  &  \RIGHTcircle  &  \RIGHTcircle  &  \CIRCLE  &  \RIGHTcircle  &  \Circle \\
Wang et al.~\cite{wang2023doasyousay}  &  \Circle  &  \Circle  &  \CIRCLE  &  \Circle  &  \Circle  &  \Circle \\
\textsc{WeJalangi}~\cite{liu2020industry}  &  \Circle  &  \Circle  &  \RIGHTcircle  &  \Circle  &  \Circle  &  \CIRCLE \\
\tool  &  \CIRCLE  &  \CIRCLE  &  \RIGHTcircle  &  \CIRCLE  &  \CIRCLE  &  \RIGHTcircle \\
\bottomrule
\end{tabular}
\begin{tablenotes}
  \item[1]{\small{Note that although \textsc{TaintMini}, \textsc{MiniTracker}, and \textsc{SPOChecker} do take event handler callbacks control flow into account, their analysis is only limited to those involving sensitive data flow between WXML and JavaScript~(e.g. <input>). Consequently, they fail to construct a complete callback control flow graph.}} 
\end{tablenotes}
\end{threeparttable}
\label{tab:related work}
\end{table}

\noindent \textbf{Privacy Analysis of Various Platforms.}
The exploration of privacy analysis in various platforms, particularly the alignment between actual app behaviors and stated privacy policies, has become increasingly pivotal in recent years. 
Considerable efforts have been directed towards assessing privacy policy adherence in mobile apps, which is often performed either automatically by analyzing sensitive API calls~\cite{slavin2016toward,zimmeck2016automated} or user inputs~\cite{wang2018guileak,nan2015uipicker}. Tools like \textsc{PoliCheck}~\cite{andow2020actions} have considered the recipients of personal data, thereby improving the accuracy of compliance analysis. Additionally, \textsc{PurPliance}~\cite{bui2021consistency} detects the data-usage purposes inconsistencies between the privacy policy and the actual behavior of Android apps. 
Beyond mobile applications, researchers have delved into other platforms. For instance, \textsc{ExtPrivA}~\cite{bui2023detection} and Ling et al.~\cite{ling2022arethey} focus on detecting inconsistencies between the privacy disclosures and data-collection behavior in browser extensions. \textsc{OVRseen}~\cite{trimananda2022ovrseen} compares network traffic and privacy policies to analyze personal data exposed by OVR apps. While previous works have provided valuable insights into privacy compliance across various platforms, our research extends these efforts to the emerging field of MiniApps. 
\section{Conclusion}
Our paper presents \tool, a tool for detecting privacy inconsistency using hybrid analysis. \tool leverages static analysis for high-level guidance in dynamic testing and cross-validates runtime behavior with privacy policies. Evaluations show \tool identifies privacy practices with an average precision, recall, and F1 score of 99.2\%, 95.6\%, and 97.4\% respectively. Our large-scale study reveals privacy inconsistency issues in the MiniApp ecosystem, with only 10,786 out of 127,460 WeChat MiniApps providing valid privacy policies, and 2,282 (21.2\%) displaying various privacy violations. Our findings have been responsibly disclosed to 2,282 developers, with 44 acknowledgments. We believe \tool will aid researchers and developers in identifying and mitigating privacy risks in MiniApps.

\section*{Acknowledgement}
This work was supported in part by the Key R\&D Program of Hubei Province (2023BAB017, 2023BAB079), the National Natural Science Foundation of China (grants No.62072046, 62302181, 62302176), HUST CSE-HongXin Joint Institute for Cyber
Security, HUST CSE-FiberHome Joint Institute for Cyber Security, and the Xiaomi Young Talents Program.

% \begin{acks}
% \end{acks}

\bibliographystyle{ACM-Reference-Format}
\bibliography{referencefixed}


\begin{thebibliography}{63}

%%% the Web address.

\ifx \showCODEN    \undefined \def \showCODEN     #1{\unskip}     \fi
\ifx \showDOI      \undefined \def \showDOI       #1{#1}\fi
\ifx \showISBNx    \undefined \def \showISBNx     #1{\unskip}     \fi
\ifx \showISBNxiii \undefined \def \showISBNxiii  #1{\unskip}     \fi
\ifx \showISSN     \undefined \def \showISSN      #1{\unskip}     \fi
\ifx \showLCCN     \undefined \def \showLCCN      #1{\unskip}     \fi
\ifx \shownote     \undefined \def \shownote      #1{#1}          \fi
\ifx \showarticletitle \undefined \def \showarticletitle #1{#1}   \fi
\ifx \showURL      \undefined \def \showURL       {\relax}        \fi
% invisible to TeX
\providecommand\bibfield[2]{#2}
\providecommand\bibinfo[2]{#2}
\providecommand\natexlab[1]{#1}
\providecommand\showeprint[2][]{arXiv:#2}

\bibitem[APP(2022)]%
        {APPI}
 \bibinfo{year}{2022}\natexlab{}.
\newblock \bibinfo{title}{Act on the Protection of Personal Information}.
\newblock \bibinfo{howpublished}{\url{https://www.ppc.go.jp/}}.
\newblock


\bibitem[CCP(2022)]%
        {CCPA}
 \bibinfo{year}{2022}\natexlab{}.
\newblock \bibinfo{title}{California Consumer Privacy Act}.
\newblock \bibinfo{howpublished}{\url{https://oag.ca.gov/privacy/ccpa}}.
\newblock


\bibitem[CPP(2022)]%
        {CPPA}
 \bibinfo{year}{2022}\natexlab{}.
\newblock \bibinfo{title}{{Consumer Privacy Protection Act}}.
\newblock \bibinfo{howpublished}{\url{https://ised-isde.canada.ca/site/innovation-better-canada/en/consumer-privacy-protection-act}}.
\newblock


\bibitem[wec(2022)]%
        {wechatshopping}
 \bibinfo{year}{2022}\natexlab{}.
\newblock \bibinfo{title}{eCommerce SaaS solution by WeChat: a complete guide}.
\newblock \bibinfo{howpublished}{\url{https://wechatwiki.com/wechat-resources/wechat-mini-shop-ecommerce-solution/}}.
\newblock


\bibitem[GDP(2022)]%
        {GDPR}
 \bibinfo{year}{2022}\natexlab{}.
\newblock \bibinfo{title}{General Data Protection Regulation}.
\newblock \bibinfo{howpublished}{\url{https://commission.europa.eu/law/law-topic/data-protection_en}}.
\newblock


\bibitem[Wec(2023a)]%
        {Wechat_risk2}
 \bibinfo{year}{2023}\natexlab{a}.
\newblock \bibinfo{title}{{First Major Analysis of WeChat Ecosystem Network Requests Finds Privacy Gaps, Undisclosed Data Sharing}}.
\newblock \bibinfo{howpublished}{\url{https://www.cpomagazine.com/data-privacy/first-major-analysis-of-wechat-ecosystem-network-requests-finds-privacy-gaps-undisclosed-data-sharing/}}.
\newblock


\bibitem[Wec(2023b)]%
        {Wechat_risk1}
 \bibinfo{year}{2023}\natexlab{b}.
\newblock \bibinfo{title}{{Should We Chat? Privacy in the WeChat Ecosystem}}.
\newblock \bibinfo{howpublished}{\url{https://citizenlab.ca/2023/06/privacy-in-the-wechat-ecosystem-full-report/}}.
\newblock


\bibitem[Wec(2023c)]%
        {Wechat_API_Documentation}
 \bibinfo{year}{2023}\natexlab{c}.
\newblock \bibinfo{title}{{WeChat API Documentation}}.
\newblock \bibinfo{howpublished}{\url{https://developers.weixin.qq.com/miniprogram/en/dev/api/}}.
\newblock


\bibitem[Wec(2023d)]%
        {Wechat_privacypolicy}
 \bibinfo{year}{2023}\natexlab{d}.
\newblock \bibinfo{title}{{WECHAT PRIVACY POLICY}}.
\newblock \bibinfo{howpublished}{\url{https://www.wechat.com/en/privacy_policy.html}}.
\newblock


\bibitem[Andow et~al\mbox{.}(2019)]%
        {andow2019policylint}
\bibfield{author}{\bibinfo{person}{Benjamin Andow}, \bibinfo{person}{Samin~Yaseer Mahmud}, \bibinfo{person}{Wenyu Wang}, \bibinfo{person}{Justin Whitaker}, \bibinfo{person}{William Enck}, \bibinfo{person}{Bradley Reaves}, \bibinfo{person}{Kapil Singh}, {and} \bibinfo{person}{Tao Xie}.} \bibinfo{year}{2019}\natexlab{}.
\newblock \showarticletitle{{P}olicy{L}int: {I}nvestigating {I}nternal {P}rivacy {P}olicy {C}ontradictions on {G}oogle {P}lay}. In \bibinfo{booktitle}{\emph{28th {USENIX} Security Symposium, {USENIX} Security 2019, Santa Clara, CA, USA, August 14-16, 2019}}, \bibfield{editor}{\bibinfo{person}{Nadia Heninger} {and} \bibinfo{person}{Patrick Traynor}} (Eds.). \bibinfo{publisher}{{USENIX} Association}, \bibinfo{pages}{585--602}.
\newblock
\urldef\tempurl%
\url{https://www.usenix.org/conference/usenixsecurity19/presentation/andow}
\showURL{%
\tempurl}


\bibitem[Andow et~al\mbox{.}(2020)]%
        {andow2020actions}
\bibfield{author}{\bibinfo{person}{Benjamin Andow}, \bibinfo{person}{Samin~Yaseer Mahmud}, \bibinfo{person}{Justin Whitaker}, \bibinfo{person}{William Enck}, \bibinfo{person}{Bradley Reaves}, \bibinfo{person}{Kapil Singh}, {and} \bibinfo{person}{Serge Egelman}.} \bibinfo{year}{2020}\natexlab{}.
\newblock \showarticletitle{{A}ctions {S}peak {L}ouder than {W}ords: {E}ntity-Sensitive {P}rivacy {P}olicy and {D}ata {F}low {A}nalysis with {P}oli{C}heck}. In \bibinfo{booktitle}{\emph{29th {USENIX} Security Symposium, {USENIX} Security 2020, August 12-14, 2020}}, \bibfield{editor}{\bibinfo{person}{Srdjan Capkun} {and} \bibinfo{person}{Franziska Roesner}} (Eds.). \bibinfo{publisher}{{USENIX} Association}, \bibinfo{pages}{985--1002}.
\newblock
\urldef\tempurl%
\url{https://www.usenix.org/conference/usenixsecurity20/presentation/andow}
\showURL{%
\tempurl}


\bibitem[Anonymous(2023)]%
        {online-form}
\bibfield{author}{\bibinfo{person}{Anonymous}.} \bibinfo{year}{2023}\natexlab{}.
\newblock \bibinfo{title}{{O}nline {D}ocumentation}.
\newblock \bibinfo{howpublished}{\url{https://docs.google.com/spreadsheets/d/1l3P7D9kIRlDiR97ndGaa8xMLXooshIaQa0peYK2kV78/edit?usp=sharing}}.
\newblock


\bibitem[{appium}(2023)]%
        {appium}
\bibfield{author}{\bibinfo{person}{{appium}}.} \bibinfo{year}{2023}\natexlab{}.
\newblock \bibinfo{title}{{appium}}.
\newblock
\newblock
\newblock
\shownote{\url{https://github.com/appium/appium}}.


\bibitem[Arzt et~al\mbox{.}(2014)]%
        {arzt2014flowdroid}
\bibfield{author}{\bibinfo{person}{Steven Arzt}, \bibinfo{person}{Siegfried Rasthofer}, \bibinfo{person}{Christian Fritz}, \bibinfo{person}{Eric Bodden}, \bibinfo{person}{Alexandre Bartel}, \bibinfo{person}{Jacques Klein}, \bibinfo{person}{Yves~Le Traon}, \bibinfo{person}{Damien Octeau}, {and} \bibinfo{person}{Patrick~D. McDaniel}.} \bibinfo{year}{2014}\natexlab{}.
\newblock \showarticletitle{{F}low{D}roid: precise context, flow, field, object-sensitive and lifecycle-aware taint analysis for {A}ndroid apps}. In \bibinfo{booktitle}{\emph{{ACM} {SIGPLAN} Conference on Programming Language Design and Implementation, {PLDI} '14, Edinburgh, United Kingdom - June 09 - 11, 2014}}, \bibfield{editor}{\bibinfo{person}{Michael F.~P. O'Boyle} {and} \bibinfo{person}{Keshav Pingali}} (Eds.). \bibinfo{publisher}{{ACM}}, \bibinfo{pages}{259--269}.
\newblock
\urldef\tempurl%
\url{https://doi.org/10.1145/2594291.2594299}
\showDOI{\tempurl}


\bibitem[Baskaran et~al\mbox{.}(2023)]%
        {baskaran2023measuring}
\bibfield{author}{\bibinfo{person}{Supraja Baskaran}, \bibinfo{person}{Lianying Zhao}, \bibinfo{person}{Mohammad Mannan}, {and} \bibinfo{person}{Amr~M. Youssef}.} \bibinfo{year}{2023}\natexlab{}.
\newblock \showarticletitle{{M}easuring the {L}eakage and {E}xploitability of {A}uthentication {S}ecrets in {S}uper-apps: {T}he {W}e{C}hat {C}ase}.
\newblock \bibinfo{journal}{\emph{CoRR}}  \bibinfo{volume}{abs/2307.09317} (\bibinfo{year}{2023}).
\newblock
\urldef\tempurl%
\url{https://doi.org/10.48550/ARXIV.2307.09317}
\showDOI{\tempurl}
\showeprint[arXiv]{2307.09317}


\bibitem[Bui et~al\mbox{.}(2023)]%
        {bui2023detection}
\bibfield{author}{\bibinfo{person}{Duc Bui}, \bibinfo{person}{Brian Tang}, {and} \bibinfo{person}{Kang~G. Shin}.} \bibinfo{year}{2023}\natexlab{}.
\newblock \showarticletitle{{D}etection of {I}nconsistencies in {P}rivacy {P}ractices of {B}rowser {E}xtensions}. In \bibinfo{booktitle}{\emph{44th {IEEE} Symposium on Security and Privacy, {SP} 2023, San Francisco, CA, USA, May 21-25, 2023}}. \bibinfo{publisher}{{IEEE}}, \bibinfo{pages}{2780--2798}.
\newblock
\urldef\tempurl%
\url{https://doi.org/10.1109/SP46215.2023.10179338}
\showDOI{\tempurl}


\bibitem[Bui et~al\mbox{.}(2021)]%
        {bui2021consistency}
\bibfield{author}{\bibinfo{person}{Duc Bui}, \bibinfo{person}{Yuan Yao}, \bibinfo{person}{Kang~G. Shin}, \bibinfo{person}{Jong{-}Min Choi}, {and} \bibinfo{person}{Junbum Shin}.} \bibinfo{year}{2021}\natexlab{}.
\newblock \showarticletitle{{C}onsistency {A}nalysis of {D}ata-Usage {P}urposes in {M}obile {A}pps}. In \bibinfo{booktitle}{\emph{{CCS} '21: 2021 {ACM} {SIGSAC} Conference on Computer and Communications Security, Virtual Event, Republic of Korea, November 15 - 19, 2021}}, \bibfield{editor}{\bibinfo{person}{Yongdae Kim}, \bibinfo{person}{Jong Kim}, \bibinfo{person}{Giovanni Vigna}, {and} \bibinfo{person}{Elaine Shi}} (Eds.). \bibinfo{publisher}{{ACM}}, \bibinfo{pages}{2824--2843}.
\newblock
\urldef\tempurl%
\url{https://doi.org/10.1145/3460120.3484536}
\showDOI{\tempurl}


\bibitem[Dong et~al\mbox{.}(2018)]%
        {dong2018frauddroid}
\bibfield{author}{\bibinfo{person}{Feng Dong}, \bibinfo{person}{Haoyu Wang}, \bibinfo{person}{Li Li}, \bibinfo{person}{Yao Guo}, \bibinfo{person}{Tegawend{\'{e}}~F. Bissyand{\'{e}}}, \bibinfo{person}{Tianming Liu}, \bibinfo{person}{Guoai Xu}, {and} \bibinfo{person}{Jacques Klein}.} \bibinfo{year}{2018}\natexlab{}.
\newblock \showarticletitle{{F}raud{D}roid: automated ad fraud detection for {A}ndroid apps}. In \bibinfo{booktitle}{\emph{Proceedings of the 2018 {ACM} Joint Meeting on European Software Engineering Conference and Symposium on the Foundations of Software Engineering, {ESEC/SIGSOFT} {FSE} 2018, Lake Buena Vista, FL, USA, November 04-09, 2018}}, \bibfield{editor}{\bibinfo{person}{Gary~T. Leavens}, \bibinfo{person}{Alessandro Garcia}, {and} \bibinfo{person}{Corina~S. Pasareanu}} (Eds.). \bibinfo{publisher}{{ACM}}, \bibinfo{pages}{257--268}.
\newblock
\urldef\tempurl%
\url{https://doi.org/10.1145/3236024.3236045}
\showDOI{\tempurl}


\bibitem[Enck et~al\mbox{.}(2010)]%
        {enck2014taintdroid}
\bibfield{author}{\bibinfo{person}{William Enck}, \bibinfo{person}{Peter Gilbert}, \bibinfo{person}{Byung{-}Gon Chun}, \bibinfo{person}{Landon~P. Cox}, \bibinfo{person}{Jaeyeon Jung}, \bibinfo{person}{Patrick~D. McDaniel}, {and} \bibinfo{person}{Anmol Sheth}.} \bibinfo{year}{2010}\natexlab{}.
\newblock \showarticletitle{{T}aint{D}roid: {A}n {I}nformation-Flow {T}racking {S}ystem for {R}ealtime {P}rivacy {M}onitoring on {S}martphones}. In \bibinfo{booktitle}{\emph{9th {USENIX} Symposium on Operating Systems Design and Implementation, {OSDI} 2010, October 4-6, 2010, Vancouver, BC, Canada, Proceedings}}, \bibfield{editor}{\bibinfo{person}{Remzi~H. Arpaci{-}Dusseau} {and} \bibinfo{person}{Brad Chen}} (Eds.). \bibinfo{publisher}{{USENIX} Association}, \bibinfo{pages}{393--407}.
\newblock
\urldef\tempurl%
\url{http://www.usenix.org/events/osdi10/tech/full\_papers/Enck.pdf}
\showURL{%
\tempurl}


\bibitem[{frida}(2023)]%
        {frida}
\bibfield{author}{\bibinfo{person}{{frida}}.} \bibinfo{year}{2023}\natexlab{}.
\newblock \bibinfo{title}{{frida}}.
\newblock
\newblock
\newblock
\shownote{\url{https://github.com/frida/frida}}.


\bibitem[Hao et~al\mbox{.}(2018)]%
        {hao2018analysis}
\bibfield{author}{\bibinfo{person}{Lei Hao}, \bibinfo{person}{Fucheng Wan}, \bibinfo{person}{Ning Ma}, {and} \bibinfo{person}{Yicheng Wang}.} \bibinfo{year}{2018}\natexlab{}.
\newblock \showarticletitle{Analysis of the Development of WeChat Mini Program}.
\newblock \bibinfo{journal}{\emph{Journal of Physics: Conference Series}} \bibinfo{volume}{1087}, \bibinfo{number}{6}, \bibinfo{pages}{062040}.
\newblock
\urldef\tempurl%
\url{https://doi.org/10.1088/1742-6596/1087/6/062040}
\showDOI{\tempurl}


\bibitem[Harkous et~al\mbox{.}(2018)]%
        {harkous2018polisis}
\bibfield{author}{\bibinfo{person}{Hamza Harkous}, \bibinfo{person}{Kassem Fawaz}, \bibinfo{person}{R{\'{e}}mi Lebret}, \bibinfo{person}{Florian Schaub}, \bibinfo{person}{Kang~G. Shin}, {and} \bibinfo{person}{Karl Aberer}.} \bibinfo{year}{2018}\natexlab{}.
\newblock \showarticletitle{{P}olisis: {A}utomated {A}nalysis and {P}resentation of {P}rivacy {P}olicies {U}sing {D}eep {L}earning}.
\newblock \bibinfo{journal}{\emph{CoRR}}  \bibinfo{volume}{abs/1802.02561} (\bibinfo{year}{2018}).
\newblock
\showeprint[arXiv]{1802.02561}
\urldef\tempurl%
\url{http://arxiv.org/abs/1802.02561}
\showURL{%
\tempurl}


\bibitem[Kumar et~al\mbox{.}(2018)]%
        {Kumar_2018}
\bibfield{author}{\bibinfo{person}{Vijay Kumar}, \bibinfo{person}{Shikha Arya}, {and} \bibinfo{person}{Vinesh~Kumar Gupta}.} \bibinfo{year}{2018}\natexlab{}.
\newblock \showarticletitle{{A}dvances in {I}ntrusion {D}etection and {P}revention {T}echniques: {A} {S}urvey}.
\newblock \bibinfo{journal}{\emph{International Journal of Computer Network and Information Security}}  \bibinfo{volume}{6} (\bibinfo{date}{Apr} \bibinfo{year}{2018}), \bibinfo{pages}{1–13}.
\newblock
\showISSN{2162-237X}
\urldef\tempurl%
\url{https://doi.org/10.5815/ijcnis.2018.06.01}
\showDOI{\tempurl}


\bibitem[Li et~al\mbox{.}(2024)]%
        {li2023minitracker}
\bibfield{author}{\bibinfo{person}{Wei Li}, \bibinfo{person}{Borui Yang}, \bibinfo{person}{Hangyu Ye}, \bibinfo{person}{Liyao Xiang}, \bibinfo{person}{Qingxiao Tao}, \bibinfo{person}{Xinbing Wang}, {and} \bibinfo{person}{Chenghu Zhou}.} \bibinfo{year}{2024}\natexlab{}.
\newblock \showarticletitle{{M}ini{T}racker: {L}arge-Scale {S}ensitive {I}nformation {T}racking in {M}ini {A}pps}.
\newblock \bibinfo{journal}{\emph{{IEEE} Trans. Dependable Secur. Comput.}} \bibinfo{volume}{21}, \bibinfo{number}{4} (\bibinfo{year}{2024}), \bibinfo{pages}{2099--2114}.
\newblock
\urldef\tempurl%
\url{https://doi.org/10.1109/TDSC.2023.3299945}
\showDOI{\tempurl}


\bibitem[Ling et~al\mbox{.}(2022)]%
        {ling2022arethey}
\bibfield{author}{\bibinfo{person}{Yuxi Ling}, \bibinfo{person}{Kailong Wang}, \bibinfo{person}{Guangdong Bai}, \bibinfo{person}{Haoyu Wang}, {and} \bibinfo{person}{Jin~Song Dong}.} \bibinfo{year}{2022}\natexlab{}.
\newblock \showarticletitle{{A}re they {T}oeing the {L}ine? {D}iagnosing {P}rivacy {C}ompliance {V}iolations among {B}rowser {E}xtensions}. In \bibinfo{booktitle}{\emph{37th {IEEE/ACM} International Conference on Automated Software Engineering, {ASE} 2022, Rochester, MI, USA, October 10-14, 2022}}. \bibinfo{publisher}{{ACM}}, \bibinfo{pages}{10:1--10:12}.
\newblock
\urldef\tempurl%
\url{https://doi.org/10.1145/3551349.3560436}
\showDOI{\tempurl}


\bibitem[Liu et~al\mbox{.}(2022)]%
        {liu2022promal}
\bibfield{author}{\bibinfo{person}{Changlin Liu}, \bibinfo{person}{Hanlin Wang}, \bibinfo{person}{Tianming Liu}, \bibinfo{person}{Diandian Gu}, \bibinfo{person}{Yun Ma}, \bibinfo{person}{Haoyu Wang}, {and} \bibinfo{person}{Xusheng Xiao}.} \bibinfo{year}{2022}\natexlab{}.
\newblock \showarticletitle{{PROMAL:} {P}recise {W}indow {T}ransition {G}raphs for {A}ndroid via {S}ynergy of {P}rogram {A}nalysis and {M}achine {L}earning}. In \bibinfo{booktitle}{\emph{44th {IEEE/ACM} 44th International Conference on Software Engineering, {ICSE} 2022, Pittsburgh, PA, USA, May 25-27, 2022}}. \bibinfo{publisher}{{ACM}}, \bibinfo{pages}{1755--1767}.
\newblock
\urldef\tempurl%
\url{https://doi.org/10.1145/3510003.3510037}
\showDOI{\tempurl}


\bibitem[Liu et~al\mbox{.}(2020)]%
        {liu2020industry}
\bibfield{author}{\bibinfo{person}{Yi Liu}, \bibinfo{person}{Jinhui Xie}, \bibinfo{person}{Jianbo Yang}, \bibinfo{person}{Shiyu Guo}, \bibinfo{person}{Yuetang Deng}, \bibinfo{person}{Shuqing Li}, \bibinfo{person}{Yechang Wu}, {and} \bibinfo{person}{Yepang Liu}.} \bibinfo{year}{2020}\natexlab{}.
\newblock \showarticletitle{{I}ndustry {P}ractice of {J}ava{S}cript {D}ynamic {A}nalysis on {W}e{C}hat {M}ini-Programs}. In \bibinfo{booktitle}{\emph{35th {IEEE/ACM} International Conference on Automated Software Engineering, {ASE} 2020, Melbourne, Australia, September 21-25, 2020}}. \bibinfo{publisher}{{IEEE}}, \bibinfo{pages}{1189--1193}.
\newblock
\urldef\tempurl%
\url{https://doi.org/10.1145/3324884.3421842}
\showDOI{\tempurl}


\bibitem[Liu et~al\mbox{.}(2023)]%
        {liu2023ex}
\bibfield{author}{\bibinfo{person}{Zhe Liu}, \bibinfo{person}{Chunyang Chen}, \bibinfo{person}{Junjie Wang}, \bibinfo{person}{Yuhui Su}, \bibinfo{person}{Yuekai Huang}, \bibinfo{person}{Jun Hu}, {and} \bibinfo{person}{Qing Wang}.} \bibinfo{year}{2023}\natexlab{}.
\newblock \showarticletitle{{E}x pede {H}erculem: {A}ugmenting {A}ctivity {T}ransition {G}raph for {A}pps via {G}raph {C}onvolution {N}etwork}. In \bibinfo{booktitle}{\emph{45th {IEEE/ACM} International Conference on Software Engineering, {ICSE} 2023, Melbourne, Australia, May 14-20, 2023}}. \bibinfo{publisher}{{IEEE}}, \bibinfo{pages}{1983--1995}.
\newblock
\urldef\tempurl%
\url{https://doi.org/10.1109/ICSE48619.2023.00168}
\showDOI{\tempurl}


\bibitem[Lu et~al\mbox{.}(2020)]%
        {lu2020demystifying}
\bibfield{author}{\bibinfo{person}{Haoran Lu}, \bibinfo{person}{Luyi Xing}, \bibinfo{person}{Yue Xiao}, \bibinfo{person}{Yifan Zhang}, \bibinfo{person}{Xiaojing Liao}, \bibinfo{person}{XiaoFeng Wang}, {and} \bibinfo{person}{Xueqiang Wang}.} \bibinfo{year}{2020}\natexlab{}.
\newblock \showarticletitle{{D}emystifying {R}esource {M}anagement {R}isks in {E}merging {M}obile {A}pp-in-App {E}cosystems}. In \bibinfo{booktitle}{\emph{{CCS} '20: 2020 {ACM} {SIGSAC} Conference on Computer and Communications Security, Virtual Event, USA, November 9-13, 2020}}, \bibfield{editor}{\bibinfo{person}{Jay Ligatti}, \bibinfo{person}{Xinming Ou}, \bibinfo{person}{Jonathan Katz}, {and} \bibinfo{person}{Giovanni Vigna}} (Eds.). \bibinfo{publisher}{{ACM}}, \bibinfo{pages}{569--585}.
\newblock
\urldef\tempurl%
\url{https://doi.org/10.1145/3372297.3417255}
\showDOI{\tempurl}


\bibitem[Meng et~al\mbox{.}(2023)]%
        {meng2023wemint}
\bibfield{author}{\bibinfo{person}{Shi Meng}, \bibinfo{person}{Liu Wang}, \bibinfo{person}{Shenao Wang}, \bibinfo{person}{Kailong Wang}, \bibinfo{person}{Xusheng Xiao}, \bibinfo{person}{Guangdong Bai}, {and} \bibinfo{person}{Haoyu Wang}.} \bibinfo{year}{2023}\natexlab{}.
\newblock \showarticletitle{{W}emint:{T}ainting {S}ensitive {D}ata {L}eaks in {W}e{C}hat {M}ini-Programs}. In \bibinfo{booktitle}{\emph{38th {IEEE/ACM} International Conference on Automated Software Engineering, {ASE} 2023, Luxembourg, September 11-15, 2023}}. \bibinfo{publisher}{{IEEE}}, \bibinfo{pages}{1403--1415}.
\newblock
\urldef\tempurl%
\url{https://doi.org/10.1109/ASE56229.2023.00151}
\showDOI{\tempurl}


\bibitem[Nan et~al\mbox{.}(2015)]%
        {nan2015uipicker}
\bibfield{author}{\bibinfo{person}{Yuhong Nan}, \bibinfo{person}{Min Yang}, \bibinfo{person}{Zhemin Yang}, \bibinfo{person}{Shunfan Zhou}, \bibinfo{person}{Guofei Gu}, {and} \bibinfo{person}{Xiaofeng Wang}.} \bibinfo{year}{2015}\natexlab{}.
\newblock \showarticletitle{{U}{I}{P}icker: {U}ser-Input {P}rivacy {I}dentification in {M}obile {A}pplications}. In \bibinfo{booktitle}{\emph{24th {USENIX} Security Symposium, {USENIX} Security 15, Washington, D.C., USA, August 12-14, 2015}}, \bibfield{editor}{\bibinfo{person}{Jaeyeon Jung} {and} \bibinfo{person}{Thorsten Holz}} (Eds.). \bibinfo{publisher}{{USENIX} Association}, \bibinfo{pages}{993--1008}.
\newblock
\urldef\tempurl%
\url{https://www.usenix.org/conference/usenixsecurity15/technical-sessions/presentation/nan}
\showURL{%
\tempurl}


\bibitem[{r3x5ur}(2023)]%
        {unveilr}
\bibfield{author}{\bibinfo{person}{{r3x5ur}}.} \bibinfo{year}{2023}\natexlab{}.
\newblock \bibinfo{title}{{unveilr}}.
\newblock
\newblock
\newblock
\shownote{\url{https://github.com/r3x5ur/unveilr}}.


\bibitem[Rao and Ko(2021)]%
        {rao2021impulsive}
\bibfield{author}{\bibinfo{person}{Qianhui Rao} {and} \bibinfo{person}{Eunju Ko}.} \bibinfo{year}{2021}\natexlab{}.
\newblock \showarticletitle{Impulsive purchasing and luxury brand loyalty in WeChat Mini Program}.
\newblock \bibinfo{journal}{\emph{Asia Pacific Journal of Marketing and Logistics}} \bibinfo{volume}{33}, \bibinfo{number}{10} (\bibinfo{year}{2021}), \bibinfo{pages}{2054--2071}.
\newblock
\showISSN{1355-5855}
\urldef\tempurl%
\url{https://doi.org/10.1108/APJML-08-2020-0621}
\showDOI{\tempurl}


\bibitem[{security-pride}(2023)]%
        {miniscope}
\bibfield{author}{\bibinfo{person}{{security-pride}}.} \bibinfo{year}{2023}\natexlab{}.
\newblock \bibinfo{title}{{MiniScope}}.
\newblock
\newblock
\newblock
\shownote{\url{https://github.com/security-pride/MiniScope}}.


\bibitem[{sensepost}(2023)]%
        {objection}
\bibfield{author}{\bibinfo{person}{{sensepost}}.} \bibinfo{year}{2023}\natexlab{}.
\newblock \bibinfo{title}{{objection}}.
\newblock
\newblock
\newblock
\shownote{\url{https://github.com/sensepost/objection}}.


\bibitem[Slavin et~al\mbox{.}(2016)]%
        {slavin2016toward}
\bibfield{author}{\bibinfo{person}{Rocky Slavin}, \bibinfo{person}{Xiaoyin Wang}, \bibinfo{person}{Mitra~Bokaei Hosseini}, \bibinfo{person}{James Hester}, \bibinfo{person}{Ram Krishnan}, \bibinfo{person}{Jaspreet Bhatia}, \bibinfo{person}{Travis~D. Breaux}, {and} \bibinfo{person}{Jianwei Niu}.} \bibinfo{year}{2016}\natexlab{}.
\newblock \showarticletitle{{T}oward a framework for detecting privacy policy violations in android application code}. In \bibinfo{booktitle}{\emph{Proceedings of the 38th International Conference on Software Engineering, {ICSE} 2016, Austin, TX, USA, May 14-22, 2016}}, \bibfield{editor}{\bibinfo{person}{Laura~K. Dillon}, \bibinfo{person}{Willem Visser}, {and} \bibinfo{person}{Laurie~A. Williams}} (Eds.). \bibinfo{publisher}{{ACM}}, \bibinfo{pages}{25--36}.
\newblock
\urldef\tempurl%
\url{https://doi.org/10.1145/2884781.2884855}
\showDOI{\tempurl}


\bibitem[Trimananda et~al\mbox{.}(2022)]%
        {trimananda2022ovrseen}
\bibfield{author}{\bibinfo{person}{Rahmadi Trimananda}, \bibinfo{person}{Hieu Le}, \bibinfo{person}{Hao Cui}, \bibinfo{person}{Janice~Tran Ho}, \bibinfo{person}{Anastasia Shuba}, {and} \bibinfo{person}{Athina Markopoulou}.} \bibinfo{year}{2022}\natexlab{}.
\newblock \showarticletitle{{O}{V}{R}seen: {A}uditing {N}etwork {T}raffic and {P}rivacy {P}olicies in {O}culus {VR}}. In \bibinfo{booktitle}{\emph{31st {USENIX} Security Symposium, {USENIX} Security 2022, Boston, MA, USA, August 10-12, 2022}}, \bibfield{editor}{\bibinfo{person}{Kevin R.~B. Butler} {and} \bibinfo{person}{Kurt Thomas}} (Eds.). \bibinfo{publisher}{{USENIX} Association}, \bibinfo{pages}{3789--3806}.
\newblock
\urldef\tempurl%
\url{https://www.usenix.org/conference/usenixsecurity22/presentation/trimananda}
\showURL{%
\tempurl}


\bibitem[{W3C}(2023a)]%
        {w3c}
\bibfield{author}{\bibinfo{person}{{W3C}}.} \bibinfo{year}{2023}\natexlab{a}.
\newblock \bibinfo{title}{{MiniApp {S}tandardization {W}hite {P}aper}}.
\newblock
\newblock
\newblock
\shownote{\url{https://www.w3.org/TR/mini-app-white-paper}}.


\bibitem[{W3C}(2023b)]%
        {subpackaging}
\bibfield{author}{\bibinfo{person}{{W3C}}.} \bibinfo{year}{2023}\natexlab{b}.
\newblock \bibinfo{title}{{MiniApp {S}ubpackaging}}.
\newblock
\newblock
\newblock
\shownote{\url{https://www.w3.org/TR/mini-app-white-paper/\#subpackaging}}.


\bibitem[Wang et~al\mbox{.}(2023a)]%
        {wang2023taintmini}
\bibfield{author}{\bibinfo{person}{Chao Wang}, \bibinfo{person}{Ronny Ko}, \bibinfo{person}{Yue Zhang}, \bibinfo{person}{Yuqing Yang}, {and} \bibinfo{person}{Zhiqiang Lin}.} \bibinfo{year}{2023}\natexlab{a}.
\newblock \showarticletitle{{T}aintmini: {D}etecting {F}low of {S}ensitive {D}ata in {M}ini-Programs with {S}tatic {T}aint {A}nalysis}. In \bibinfo{booktitle}{\emph{45th {IEEE/ACM} International Conference on Software Engineering, {ICSE} 2023, Melbourne, Australia, May 14-20, 2023}}. \bibinfo{publisher}{{IEEE}}, \bibinfo{pages}{932--944}.
\newblock
\urldef\tempurl%
\url{https://doi.org/10.1109/ICSE48619.2023.00086}
\showDOI{\tempurl}


\bibitem[Wang et~al\mbox{.}(2023b)]%
        {wang2023uncovering}
\bibfield{author}{\bibinfo{person}{Chao Wang}, \bibinfo{person}{Yue Zhang}, {and} \bibinfo{person}{Zhiqiang Lin}.} \bibinfo{year}{2023}\natexlab{b}.
\newblock \showarticletitle{{U}ncovering and {E}xploiting {H}idden {A}{P}{I}s in {M}obile {S}uper {A}pps}.
\newblock \bibinfo{journal}{\emph{CoRR}}  \bibinfo{volume}{abs/2306.08134} (\bibinfo{year}{2023}).
\newblock
\urldef\tempurl%
\url{https://doi.org/10.48550/ARXIV.2306.08134}
\showDOI{\tempurl}
\showeprint[arXiv]{2306.08134}


\bibitem[Wang et~al\mbox{.}(2023c)]%
        {wang2023sats}
\bibfield{author}{\bibinfo{person}{Shenao Wang}, \bibinfo{person}{Yanjie Zhao}, \bibinfo{person}{Kailong Wang}, {and} \bibinfo{person}{Haoyu Wang}.} \bibinfo{year}{2023}\natexlab{c}.
\newblock \showarticletitle{{O}n the {U}sage-scenario-based {D}ata {M}inimization in {M}ini {P}rograms}. In \bibinfo{booktitle}{\emph{Proceedings of the 2023 {ACM} Workshop on Secure and Trustworthy Superapps, SaTS 2023, Copenhagen, Denmark, 26 November 2023}}, \bibfield{editor}{\bibinfo{person}{Zhiqiang Lin} {and} \bibinfo{person}{Xiaojing Liao}} (Eds.). \bibinfo{publisher}{{ACM}}, \bibinfo{pages}{29--32}.
\newblock
\urldef\tempurl%
\url{https://doi.org/10.1145/3605762.3624435}
\showDOI{\tempurl}


\bibitem[Wang et~al\mbox{.}(2022)]%
        {wang2022characterizing}
\bibfield{author}{\bibinfo{person}{Tao Wang}, \bibinfo{person}{Qingxin Xu}, \bibinfo{person}{Xiaoning Chang}, \bibinfo{person}{Wensheng Dou}, \bibinfo{person}{Jiaxin Zhu}, \bibinfo{person}{Jinhui Xie}, \bibinfo{person}{Yuetang Deng}, \bibinfo{person}{Jianbo Yang}, \bibinfo{person}{Jiaheng Yang}, \bibinfo{person}{Jun Wei}, {and} \bibinfo{person}{Tao Huang}.} \bibinfo{year}{2022}\natexlab{}.
\newblock \showarticletitle{{C}haracterizing and {D}etecting {B}ugs in {W}e{C}hat {M}ini-Programs}. In \bibinfo{booktitle}{\emph{44th {IEEE/ACM} 44th International Conference on Software Engineering, {ICSE} 2022, Pittsburgh, PA, USA, May 25-27, 2022}}. \bibinfo{publisher}{{ACM}}, \bibinfo{pages}{363--375}.
\newblock
\urldef\tempurl%
\url{https://doi.org/10.1145/3510003.3510114}
\showDOI{\tempurl}


\bibitem[Wang et~al\mbox{.}(2018)]%
        {wang2018guileak}
\bibfield{author}{\bibinfo{person}{Xiaoyin Wang}, \bibinfo{person}{Xue Qin}, \bibinfo{person}{Mitra~Bokaei Hosseini}, \bibinfo{person}{Rocky Slavin}, \bibinfo{person}{Travis~D. Breaux}, {and} \bibinfo{person}{Jianwei Niu}.} \bibinfo{year}{2018}\natexlab{}.
\newblock \showarticletitle{{G}{U}{I}{L}eak: tracing privacy policy claims on user input data for {A}ndroid applications}. In \bibinfo{booktitle}{\emph{Proceedings of the 40th International Conference on Software Engineering, {ICSE} 2018, Gothenburg, Sweden, May 27 - June 03, 2018}}, \bibfield{editor}{\bibinfo{person}{Michel Chaudron}, \bibinfo{person}{Ivica Crnkovic}, \bibinfo{person}{Marsha Chechik}, {and} \bibinfo{person}{Mark Harman}} (Eds.). \bibinfo{publisher}{{ACM}}, \bibinfo{pages}{37--47}.
\newblock
\urldef\tempurl%
\url{https://doi.org/10.1145/3180155.3180196}
\showDOI{\tempurl}


\bibitem[Wang et~al\mbox{.}(2024a)]%
        {wang2023doasyousay}
\bibfield{author}{\bibinfo{person}{Yin Wang}, \bibinfo{person}{Ming Fan}, \bibinfo{person}{Junfeng Liu}, \bibinfo{person}{Junjie Tao}, \bibinfo{person}{Wuxia Jin}, \bibinfo{person}{Haijun Wang}, \bibinfo{person}{Qi Xiong}, {and} \bibinfo{person}{Ting Liu}.} \bibinfo{year}{2024}\natexlab{a}.
\newblock \showarticletitle{Do as You Say: Consistency Detection of Data Practice in Program Code and Privacy Policy in Mini-App}.
\newblock \bibinfo{journal}{\emph{IEEE Transactions on Software Engineering}} (\bibinfo{year}{2024}), \bibinfo{pages}{1--23}.
\newblock
\urldef\tempurl%
\url{https://doi.org/10.1109/TSE.2024.3479288}
\showDOI{\tempurl}


\bibitem[Wang et~al\mbox{.}(2024b)]%
        {wang2024minichecker}
\bibfield{author}{\bibinfo{person}{Yin Wang}, \bibinfo{person}{Ming Fan}, \bibinfo{person}{Hao Zhou}, \bibinfo{person}{Haijun Wang}, \bibinfo{person}{Wuxia Jin}, \bibinfo{person}{Jiajia Li}, \bibinfo{person}{Wenbo Chen}, \bibinfo{person}{Shijie Li}, \bibinfo{person}{Yu Zhang}, \bibinfo{person}{Deqiang Han}, {and} \bibinfo{person}{Ting Liu}.} \bibinfo{year}{2024}\natexlab{b}.
\newblock \showarticletitle{MiniChecker: Detecting Data Privacy Risk of Abusive Permission Request Behavior in Mini-Programs}. In \bibinfo{booktitle}{\emph{Proceedings of the 39th IEEE/ACM International Conference on Automated Software Engineering}} (Sacramento, CA, USA) \emph{(\bibinfo{series}{ASE '24})}. \bibinfo{publisher}{Association for Computing Machinery}, \bibinfo{address}{New York, NY, USA}, \bibinfo{pages}{1667–1679}.
\newblock
\showISBNx{9798400712487}
\urldef\tempurl%
\url{https://doi.org/10.1145/3691620.3695534}
\showDOI{\tempurl}


\bibitem[{xdmjun}(2023)]%
        {wxappUnpacker}
\bibfield{author}{\bibinfo{person}{{xdmjun}}.} \bibinfo{year}{2023}\natexlab{}.
\newblock \bibinfo{title}{{wxappUnpacker}}.
\newblock
\newblock
\newblock
\shownote{\url{https://github.com/xdmjun/wxappUnpacker}}.


\bibitem[Yamaguchi et~al\mbox{.}(2014)]%
        {yamaguchi2014modeling}
\bibfield{author}{\bibinfo{person}{Fabian Yamaguchi}, \bibinfo{person}{Nico Golde}, \bibinfo{person}{Daniel Arp}, {and} \bibinfo{person}{Konrad Rieck}.} \bibinfo{year}{2014}\natexlab{}.
\newblock \showarticletitle{{M}odeling and {D}iscovering {V}ulnerabilities with {C}ode {P}roperty {G}raphs}. In \bibinfo{booktitle}{\emph{2014 {IEEE} Symposium on Security and Privacy, {SP} 2014, Berkeley, CA, USA, May 18-21, 2014}}. \bibinfo{publisher}{{IEEE} Computer Society}, \bibinfo{pages}{590--604}.
\newblock
\urldef\tempurl%
\url{https://doi.org/10.1109/SP.2014.44}
\showDOI{\tempurl}


\bibitem[Yan et~al\mbox{.}(2023)]%
        {yan2023muid}
\bibfield{author}{\bibinfo{person}{Ziqiang Yan}, \bibinfo{person}{Ming Fan}, \bibinfo{person}{Yin Wang}, \bibinfo{person}{Jifei Shi}, \bibinfo{person}{Haoran Wang}, {and} \bibinfo{person}{Ting Liu}.} \bibinfo{year}{2023}\natexlab{}.
\newblock \showarticletitle{MUID: Detecting Sensitive User Inputs in Miniapp Ecosystems}. In \bibinfo{booktitle}{\emph{Proceedings of the 2023 ACM Workshop on Secure and Trustworthy Superapps}} (Copenhagen, Denmark) \emph{(\bibinfo{series}{SaTS '23})}. \bibinfo{publisher}{Association for Computing Machinery}, \bibinfo{address}{New York, NY, USA}, \bibinfo{pages}{17–21}.
\newblock
\showISBNx{9798400702587}
\urldef\tempurl%
\url{https://doi.org/10.1145/3605762.3624429}
\showDOI{\tempurl}


\bibitem[Yang et~al\mbox{.}(2018)]%
        {yang2018wtg}
\bibfield{author}{\bibinfo{person}{Shengqian Yang}, \bibinfo{person}{Haowei Wu}, \bibinfo{person}{Hailong Zhang}, \bibinfo{person}{Yan Wang}, \bibinfo{person}{Chandrasekar Swaminathan}, \bibinfo{person}{Dacong Yan}, {and} \bibinfo{person}{Atanas Rountev}.} \bibinfo{year}{2018}\natexlab{}.
\newblock \showarticletitle{{S}tatic window transition graphs for {A}ndroid}.
\newblock \bibinfo{journal}{\emph{Autom. Softw. Eng.}} \bibinfo{volume}{25}, \bibinfo{number}{4} (\bibinfo{year}{2018}), \bibinfo{pages}{833--873}.
\newblock
\urldef\tempurl%
\url{https://doi.org/10.1007/S10515-018-0237-6}
\showDOI{\tempurl}


\bibitem[Yang et~al\mbox{.}(2015)]%
        {yang2015ccfg}
\bibfield{author}{\bibinfo{person}{Shengqian Yang}, \bibinfo{person}{Dacong Yan}, \bibinfo{person}{Haowei Wu}, \bibinfo{person}{Yan Wang}, {and} \bibinfo{person}{Atanas Rountev}.} \bibinfo{year}{2015}\natexlab{}.
\newblock \showarticletitle{{S}tatic {C}ontrol-Flow {A}nalysis of {U}ser-Driven {C}allbacks in {A}ndroid {A}pplications}. In \bibinfo{booktitle}{\emph{37th {IEEE/ACM} International Conference on Software Engineering, {ICSE} 2015, Florence, Italy, May 16-24, 2015, Volume 1}}, \bibfield{editor}{\bibinfo{person}{Antonia Bertolino}, \bibinfo{person}{Gerardo Canfora}, {and} \bibinfo{person}{Sebastian~G. Elbaum}} (Eds.). \bibinfo{publisher}{{IEEE} Computer Society}, \bibinfo{pages}{89--99}.
\newblock
\urldef\tempurl%
\url{https://doi.org/10.1109/ICSE.2015.31}
\showDOI{\tempurl}


\bibitem[Yang et~al\mbox{.}(2022a)]%
        {yang2022permdroid}
\bibfield{author}{\bibinfo{person}{Shuaihao Yang}, \bibinfo{person}{Zigang Zeng}, {and} \bibinfo{person}{Wei Song}.} \bibinfo{year}{2022}\natexlab{a}.
\newblock \showarticletitle{{P}erm{D}roid: automatically testing permission-related behaviour of {A}ndroid applications}. In \bibinfo{booktitle}{\emph{{ISSTA} '22: 31st {ACM} {SIGSOFT} International Symposium on Software Testing and Analysis, Virtual Event, South Korea, July 18 - 22, 2022}}, \bibfield{editor}{\bibinfo{person}{Sukyoung Ryu} {and} \bibinfo{person}{Yannis Smaragdakis}} (Eds.). \bibinfo{publisher}{{ACM}}, \bibinfo{pages}{593--604}.
\newblock
\urldef\tempurl%
\url{https://doi.org/10.1145/3533767.3534221}
\showDOI{\tempurl}


\bibitem[Yang et~al\mbox{.}(2023)]%
        {yang2023sok}
\bibfield{author}{\bibinfo{person}{Yuqing Yang}, \bibinfo{person}{Chao Wang}, \bibinfo{person}{Yue Zhang}, {and} \bibinfo{person}{Zhiqiang Lin}.} \bibinfo{year}{2023}\natexlab{}.
\newblock \showarticletitle{{S}o{K}: {D}ecoding the {S}uper {A}pp {E}nigma: {T}he {S}ecurity {M}echanisms, {T}hreats, and {T}rade-offs in {O}{S}-alike {A}pps}.
\newblock \bibinfo{journal}{\emph{CoRR}}  \bibinfo{volume}{abs/2306.07495} (\bibinfo{year}{2023}).
\newblock
\urldef\tempurl%
\url{https://doi.org/10.48550/ARXIV.2306.07495}
\showDOI{\tempurl}
\showeprint[arXiv]{2306.07495}


\bibitem[Yang et~al\mbox{.}(2022b)]%
        {yang2022cross}
\bibfield{author}{\bibinfo{person}{Yuqing Yang}, \bibinfo{person}{Yue Zhang}, {and} \bibinfo{person}{Zhiqiang Lin}.} \bibinfo{year}{2022}\natexlab{b}.
\newblock \showarticletitle{{C}ross {M}iniapp {R}equest {F}orgery: {R}oot {C}auses, {A}ttacks, and {V}ulnerability {D}etection}. In \bibinfo{booktitle}{\emph{Proceedings of the 2022 {ACM} {SIGSAC} Conference on Computer and Communications Security, {CCS} 2022, Los Angeles, CA, USA, November 7-11, 2022}}, \bibfield{editor}{\bibinfo{person}{Heng Yin}, \bibinfo{person}{Angelos Stavrou}, \bibinfo{person}{Cas Cremers}, {and} \bibinfo{person}{Elaine Shi}} (Eds.). \bibinfo{publisher}{{ACM}}, \bibinfo{pages}{3079--3092}.
\newblock
\urldef\tempurl%
\url{https://doi.org/10.1145/3548606.3560597}
\showDOI{\tempurl}


\bibitem[Zhang et~al\mbox{.}(2022a)]%
        {zhang2023small}
\bibfield{author}{\bibinfo{person}{Jianyi Zhang}, \bibinfo{person}{Leixin Yang}, \bibinfo{person}{Yuyang Han}, \bibinfo{person}{Zhi Sun}, {and} \bibinfo{person}{Zixiao Xiang}.} \bibinfo{year}{2022}\natexlab{a}.
\newblock \showarticletitle{{A} {S}mall {L}eak {W}ill {S}ink {M}any {S}hips: {V}ulnerabilities {R}elated to {M}ini {P}rograms {P}ermissions}.
\newblock \bibinfo{journal}{\emph{CoRR}}  \bibinfo{volume}{abs/2205.15202} (\bibinfo{year}{2022}).
\newblock
\urldef\tempurl%
\url{https://doi.org/10.48550/ARXIV.2205.15202}
\showDOI{\tempurl}
\showeprint[arXiv]{2205.15202}


\bibitem[Zhang et~al\mbox{.}(2022b)]%
        {zhang2022identity}
\bibfield{author}{\bibinfo{person}{Lei Zhang}, \bibinfo{person}{Zhibo Zhang}, \bibinfo{person}{Ancong Liu}, \bibinfo{person}{Yinzhi Cao}, \bibinfo{person}{Xiaohan Zhang}, \bibinfo{person}{Yanjun Chen}, \bibinfo{person}{Yuan Zhang}, \bibinfo{person}{Guangliang Yang}, {and} \bibinfo{person}{Min Yang}.} \bibinfo{year}{2022}\natexlab{b}.
\newblock \showarticletitle{{I}dentity {C}onfusion in {W}eb{V}iew-based {M}obile {A}pp-in-app {E}cosystems}. In \bibinfo{booktitle}{\emph{31st {USENIX} Security Symposium, {USENIX} Security 2022, Boston, MA, USA, August 10-12, 2022}}, \bibfield{editor}{\bibinfo{person}{Kevin R.~B. Butler} {and} \bibinfo{person}{Kurt Thomas}} (Eds.). \bibinfo{publisher}{{USENIX} Association}, \bibinfo{pages}{1597--1613}.
\newblock
\urldef\tempurl%
\url{https://www.usenix.org/conference/usenixsecurity22/presentation/zhang-lei}
\showURL{%
\tempurl}


\bibitem[Zhang et~al\mbox{.}(2023a)]%
        {zhang2023spochecker}
\bibfield{author}{\bibinfo{person}{Xiaohan Zhang}, \bibinfo{person}{Yang Wang}, \bibinfo{person}{Xin Zhang}, \bibinfo{person}{Ziqi Huang}, \bibinfo{person}{Lei Zhang}, {and} \bibinfo{person}{Min Yang}.} \bibinfo{year}{2023}\natexlab{a}.
\newblock \showarticletitle{{U}nderstanding {P}rivacy {O}ver-collection in {W}e{C}hat {S}ub-app {E}cosystem}.
\newblock \bibinfo{journal}{\emph{CoRR}}  \bibinfo{volume}{abs/2306.08391} (\bibinfo{year}{2023}).
\newblock
\urldef\tempurl%
\url{https://doi.org/10.48550/ARXIV.2306.08391}
\showDOI{\tempurl}
\showeprint[arXiv]{2306.08391}


\bibitem[Zhang et~al\mbox{.}(2021)]%
        {zhang2021measurement}
\bibfield{author}{\bibinfo{person}{Yue Zhang}, \bibinfo{person}{Bayan Turkistani}, \bibinfo{person}{Allen~Yuqing Yang}, \bibinfo{person}{Chaoshun Zuo}, {and} \bibinfo{person}{Zhiqiang Lin}.} \bibinfo{year}{2021}\natexlab{}.
\newblock \showarticletitle{{A} {M}easurement {S}tudy of {W}echat {M}ini-Apps}. In \bibinfo{booktitle}{\emph{{SIGMETRICS} '21: {ACM} {SIGMETRICS} / International Conference on Measurement and Modeling of Computer Systems, Virtual Event, China, June 14-18, 2021}}, \bibfield{editor}{\bibinfo{person}{Longbo Huang}, \bibinfo{person}{Anshul Gandhi}, \bibinfo{person}{Negar Kiyavash}, {and} \bibinfo{person}{Jia Wang}} (Eds.). \bibinfo{publisher}{{ACM}}, \bibinfo{pages}{19--20}.
\newblock
\urldef\tempurl%
\url{https://doi.org/10.1145/3410220.3460106}
\showDOI{\tempurl}


\bibitem[Zhang et~al\mbox{.}(2023b)]%
        {zhang2023dont}
\bibfield{author}{\bibinfo{person}{Yue Zhang}, \bibinfo{person}{Yuqing Yang}, {and} \bibinfo{person}{Zhiqiang Lin}.} \bibinfo{year}{2023}\natexlab{b}.
\newblock \showarticletitle{{D}on't {L}eak {Y}our {K}eys: {U}nderstanding, {M}easuring, and {E}xploiting the {A}pp{S}ecret {L}eaks in {M}ini-Programs}.
\newblock \bibinfo{journal}{\emph{CoRR}}  \bibinfo{volume}{abs/2306.08151} (\bibinfo{year}{2023}).
\newblock
\urldef\tempurl%
\url{https://doi.org/10.48550/ARXIV.2306.08151}
\showDOI{\tempurl}
\showeprint[arXiv]{2306.08151}


\bibitem[Zhang et~al\mbox{.}(2024)]%
        {zhang2024minicat}
\bibfield{author}{\bibinfo{person}{Zidong Zhang}, \bibinfo{person}{Qingsheng Hou}, \bibinfo{person}{Lingyun Ying}, \bibinfo{person}{Wenrui Diao}, \bibinfo{person}{Yacong Gu}, \bibinfo{person}{Rui Li}, \bibinfo{person}{Shanqing Guo}, {and} \bibinfo{person}{Haixin Duan}.} \bibinfo{year}{2024}\natexlab{}.
\newblock \showarticletitle{MiniCAT: Understanding and Detecting Cross-Page Request Forgery Vulnerabilities in Mini-Programs}. In \bibinfo{booktitle}{\emph{Proceedings of the 2024 ACM SIGSAC Conference on Computer and Communications Security, Salt Lake City, UT, USA}}.
\newblock
\urldef\tempurl%
\url{https://doi.org/10.1145/3658644.3670294}
\showDOI{\tempurl}


\bibitem[Zhao et~al\mbox{.}(2022)]%
        {zhao2022ca4p483}
\bibfield{author}{\bibinfo{person}{Kaifa Zhao}, \bibinfo{person}{Le Yu}, \bibinfo{person}{Shiyao Zhou}, \bibinfo{person}{Jing Li}, \bibinfo{person}{Xiapu Luo}, \bibinfo{person}{Yat Fei~Aemon Chiu}, {and} \bibinfo{person}{Yutong Liu}.} \bibinfo{year}{2022}\natexlab{}.
\newblock \showarticletitle{{A} {F}ine-grained {C}hinese {S}oftware {P}rivacy {P}olicy {D}ataset for {S}equence {L}abeling and {R}egulation {C}ompliant {I}dentification}.
\newblock \bibinfo{journal}{\emph{CoRR}}  \bibinfo{volume}{abs/2212.04357} (\bibinfo{year}{2022}).
\newblock
\urldef\tempurl%
\url{https://doi.org/10.48550/ARXIV.2212.04357}
\showDOI{\tempurl}
\showeprint[arXiv]{2212.04357}


\bibitem[Zhao et~al\mbox{.}(2023)]%
        {zhao2023signature}
\bibfield{author}{\bibinfo{person}{Yanjie Zhao}, \bibinfo{person}{Yue Zhang}, {and} \bibinfo{person}{Haoyu Wang}.} \bibinfo{year}{2023}\natexlab{}.
\newblock \showarticletitle{Potential Risks Arising from the Absence of Signature Verification in Miniapp Plugins}. In \bibinfo{booktitle}{\emph{Proceedings of the 2023 ACM Workshop on Secure and Trustworthy Superapps}} (Copenhagen, Denmark) \emph{(\bibinfo{series}{SaTS '23})}. \bibinfo{publisher}{Association for Computing Machinery}, \bibinfo{address}{New York, NY, USA}, \bibinfo{pages}{59–64}.
\newblock
\showISBNx{9798400702587}
\urldef\tempurl%
\url{https://doi.org/10.1145/3605762.3624433}
\showDOI{\tempurl}


\bibitem[Zimmeck et~al\mbox{.}(2016)]%
        {zimmeck2016automated}
\bibfield{author}{\bibinfo{person}{Sebastian Zimmeck}, \bibinfo{person}{Ziqi Wang}, \bibinfo{person}{Lieyong Zou}, \bibinfo{person}{Roger Iyengar}, \bibinfo{person}{Bin Liu}, \bibinfo{person}{Florian Schaub}, \bibinfo{person}{Shomir Wilson}, \bibinfo{person}{Norman~M. Sadeh}, \bibinfo{person}{Steven~M. Bellovin}, {and} \bibinfo{person}{Joel~R. Reidenberg}.} \bibinfo{year}{2016}\natexlab{}.
\newblock \showarticletitle{{A}utomated {A}nalysis of {P}rivacy {R}equirements for {M}obile {A}pps}. In \bibinfo{booktitle}{\emph{2016 {AAAI} Fall Symposia, Arlington, Virginia, USA, November 17-19, 2016}}. \bibinfo{publisher}{{AAAI} Press}.
\newblock
\urldef\tempurl%
\url{http://aaai.org/ocs/index.php/FSS/FSS16/paper/view/14113}
\showURL{%
\tempurl}


\end{thebibliography}

\end{document}